\documentstyle[aps,epsf]{revtex}

\begin{document}
\newcommand{\fslash}[1]{ \not \! #1}
\newcommand{\Cf}{\frac{4}{3}}
\newcommand{\intk}{\int \frac{d^4k}{(2\pi)^4}}
\newcommand{\intl}{\int \frac{d^4l}{(2\pi)^4}}
\newcommand{\ndrl}{\mu^\epsilon \int \frac{d^dl}{(2\pi)^d}}
\newcommand{\Gtilde}{ \tilde{\Gamma} }
\newcommand{\half}{\frac{1}{2}}
\newcommand{\dr}{\stackrel{\rightarrow}{\nabla}}
\newcommand{\dl}{\stackrel{\leftarrow}{\nabla}}
\newcommand{\MSb}{$\overline{MS}$ }
\newcommand{\pom}{$\frac{\vec{p}}{M}$}
\newcommand{\as}{\alpha_s} 
\newcommand{\oneloop}{^{[1]}} 
\newcommand{\soneloop}{{[1]}} 

\preprint{
\begin{minipage}{4.5cm}
GUTPA-99-09-01
\end{minipage}}
\author{Peter Boyle \\ Christine Davies}
\address{Department of Physics and Astronomy, University of Glasgow, Glasgow G12 8QQ, UK}
\author{UKQCD Collaboration}

\title{One loop renormalisation of Lattice NRQCD currents for semileptonic $B\rightarrow D^{(\ast)}$ decays
to order \pom}
\maketitle

\abstract{
We present the results of a perturbative calculation to 
match the axial and vector currents for
semileptonic $B\rightarrow D^{(\ast)}$ decays
in lattice NRQCD to the continuum \MSb scheme. 
The matching is performed to $O(\alpha_s\frac{\vec{p}}{M})$
in Feynman gauge and in the on-shell renormalisation scheme.

The spatial and temporal components of the currents renormalise differently;
to this order the matching
involves a straightforward renormalisation for the $V_0$
and $A_k$ currents, and a rank two and four mixing matrix
for the $A_0$ and $V_k$ currents respectively.
The resultant one loop corrections are of $O(5\%)$,
boding well for the accuracy of forthcoming simulations.
}

\section{Introduction}
\label{SecIntro}
Forthcoming $B$ physics experiments such as BaBar, Belle, CLEO III and  LHC-b 
will greatly improve the accuracy with which we know the rates of different decay modes
of the B meson, allowing improved determinations of the CKM matrix elements
involving the $b$ quark. The exclusive decay 
$\bar{B}\rightarrow D l \bar{\nu}$ gives access to $V_{cb}$.
For example, at BaBar \cite{BaBarBook} the decay amplitude at zero recoil
will be determined with a statistical error of less than 2\% 
and with similar level of systematic errors, a factor of three 
improvement over CLEO and LEP data.
In order to extract $V_{cb}$ we must divide out the 
form factor at zero recoil, introducing associated theoretical uncertainties.
The form factors for $B\rightarrow D$ decays are calculable
within lattice QCD \footnote{ There are a number of recent reviews of the status of lattice
calculations \cite{KronfeldReview}}, and it will 
be highly challenging for the community to produce
model independent predictions of the form factors 
with sufficient statistical and systematic precision
that theoretical uncertainty matches the improved experimental
data.

The decay amplitudes can be factorised into 
hadronic form factors and perturbative electroweak terms
as follows:

        \begin{equation}
        M(B\rightarrow D^{(\ast)} l^- \bar{\nu}) = 
        -\frac{i G_F}{ \sqrt{2} } V_{cb}
        \bar{u}_l \gamma_\mu ( 1-\gamma_5 ) v_\nu H^{\mu^{(\ast)}},
        \end{equation}
where 
        \begin{equation}
        H^\mu = \langle D(p^\prime) | V^\mu | B(p) \rangle = 
        f_+(q^2) ( p+p^\prime )^\mu + f_-(q^2) (p-p^\prime)^\mu ,
        \end{equation}
        \begin{equation}
        H^{\mu^\ast}=\langle D^{\ast}(p^\prime,\epsilon) | 
                          V^\mu-A^\mu | B(p) \rangle =
        \left\{
        \begin{array}{c}
        \frac{2 i \epsilon^{\mu\nu\alpha\beta}}{M_B+M_{D^\ast}} 
        \epsilon_\nu^\ast p^\prime_\alpha p_\beta V(q^2) \\
         - (M_B+M_{D^\ast}) \epsilon^{\ast\mu} A_1(q^2)\\
         + \frac{ \epsilon^\ast \cdot q}{M_B+M_{D^\ast}} 
        (p+p^\prime)^\mu A_2(q^2) 
        \end{array}\right\}.
        \end{equation}

These are often reparametrised in terms of form factors in the heavy quark effective theory,

        \begin{equation}
        \langle D(p^\prime) | V^\mu | B(p) \rangle = \sqrt{M_B M_D} \left\{
        h_+(v\cdot v^\prime) ( v+v^\prime )^\mu + h_-(v\cdot v^\prime) (v-v^\prime)^\mu  \right\}
        \end{equation}
        \begin{equation}
        \langle D^{\ast}(p^\prime,\epsilon) | 
                          V^\mu-A^\mu | B(p) \rangle = \sqrt{M_B M_{D^\ast}}\left\{
        \begin{array}{c}
        i \epsilon^{\mu\nu\alpha\beta}
        \epsilon_\nu^\ast v^\prime_\alpha v_\beta h_V(v\cdot v^\prime) \\
         - \epsilon^{\ast\mu} (v\cdot v^\prime+1) h_{A_1}(v\cdot v^\prime)\\
         + \epsilon^\ast \cdot v v^\mu h_{A_2}(v\cdot v^\prime) \\
         + \epsilon^\ast \cdot v v^{\prime\mu} h_{A_3}(v\cdot v^\prime)
        \end{array}\right\}
        \end{equation}

The weak (vector and axial) currents in a lattice-regulated 
theory differ from the continuum \MSb currents via
their different ultra-violet cut-offs. 
A correction between the results in the two regularisation
schemes is therefore required in order to use simulated
lattice hadronic currents in the more commonly quoted continuum \MSb scheme. 
Lattice QCD and \MSb have the same infra-red behaviour, so that these
correction factors should be dominated by the
differing ultra-violet behaviour and one expects that these
corrections are calculable within perturbation theory.
Knowledge of these renormalisation constants enables the 
various elements of the Cabibbo-Kobayashi-Maskawa weak mixing
matrix to be probed using experimental decay rates and lattice calculations of
hadronic matrix elements in conjunction with standard electroweak perturbation theory.

This paper aims to make possible
the accurate determination of $B\rightarrow D$ matrix elements
by showing how to construct the \MSb currents when
using the $O(\frac{1}{M})$ lattice NRQCD 
action to simulate both the $b$ and $c$ quarks.
We perturbatively match the corrections to the weak currents
between on-shell quark states in lattice NRQCD to the same
current in continuum \MSb with fully relativistic quarks, working consistently
to $O(\frac{1}{M})$ throughout.
The paper will be constructed as follows: in Section~\ref{SecContinuum}
we will present the one loop correction to the continuum current
evaluated between an incoming $b$ quark and outgoing $c$ quark, retaining
all terms to order $\frac{\vec{p}}{M}$. 
This is matched to continuum Euclidean
NRQCD to $O(\frac{1}{M})$, giving a basis of continuum NRQCD operators
for each of the currents $V_0$, $V_k$, $A_0$, $A_k$.
In Section~\ref{SecLattice} we select a basis of operators in the
Lattice theory and define the one-loop mixing matrix, 
giving the
projection of each element of a basis for a given Lattice current at one loop
on the basis of continuum NRQCD operators.
Finally, we use these results to construct relativistic \MSb currents
to this order from lattice operators in Section~\ref{SecMatching}. 

\section{Continuum Calculation}
\label{SecContinuum}
\begin{figure}[hbt]
\caption{
\label{ContVert} 
$\alpha_s \Lambda^{\Gamma^{[1]}}$ : Continuum vertex correction to 
$b\rightarrow c$ scattering diagram via weak current $\Gamma = \gamma_\mu$ or 
$\Gamma = \gamma_\mu \gamma_5$.
}
\setlength\epsfxsize{200pt} \setlength\epsfysize{150pt}
\epsfbox{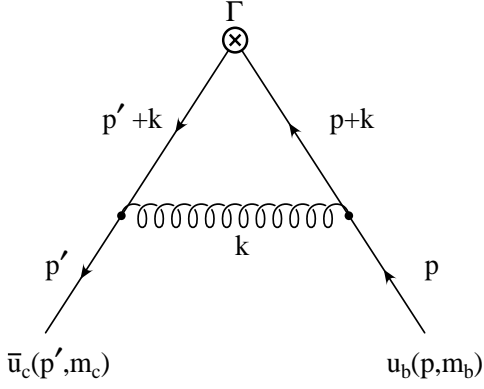}
\end{figure}

\begin{figure}[hbt]
\caption{
\label{ContLeg1} 
$\half \as \Gamma Z_{\psi_b}^{[1]}$ : First leg correction to $b\rightarrow c$ 
scattering diagram via weak current $\Gamma = \gamma_\mu$ or 
$\Gamma = \gamma_\mu \gamma_5$.
}
\setlength\epsfxsize{200pt} \setlength\epsfysize{150pt}
\epsfbox{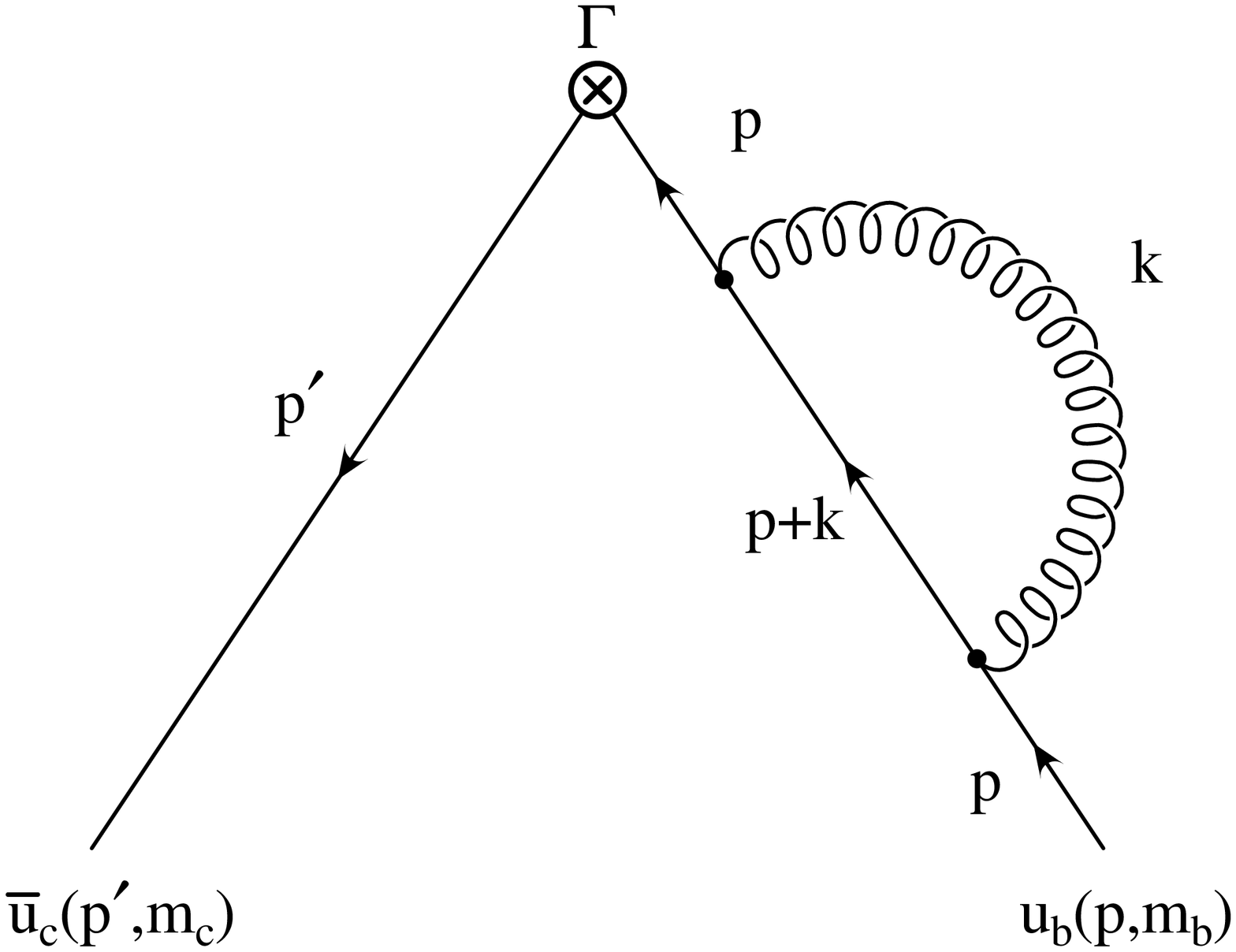}
\end{figure}

\begin{figure}[hbt]
\caption{
\label{ContLeg2} 
$\half \as \Gamma Z_{\psi_c}^{[1]}$ : Second leg correction to $b\rightarrow c$ 
scattering diagram via weak current  $\Gamma = \gamma_\mu$ or 
$\Gamma = \gamma_\mu \gamma_5$.
}
\setlength\epsfxsize{200pt} \setlength\epsfysize{150pt}
\epsfbox{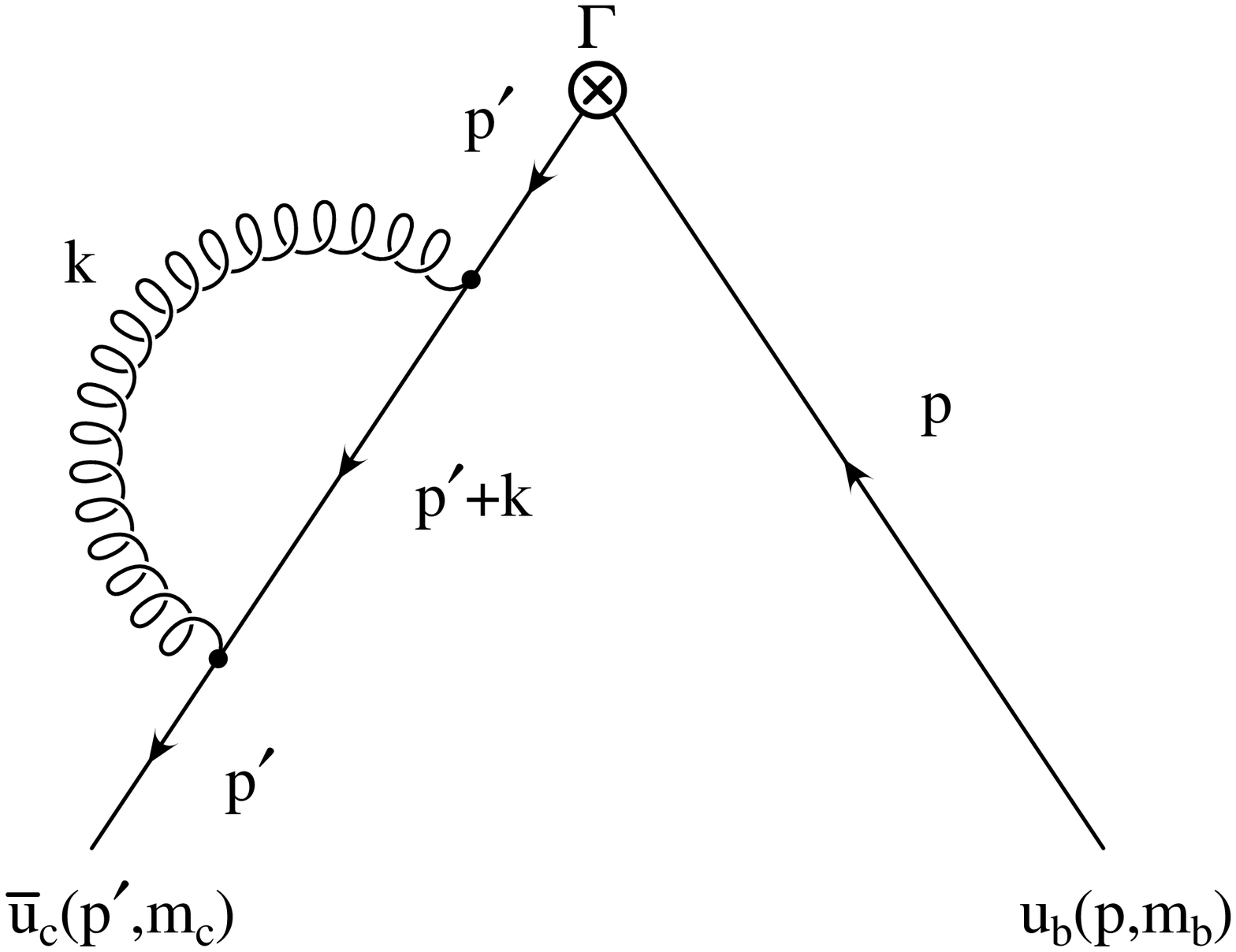}
\end{figure}

We require the continuum renormalisation of the Vector and Axial currents 
involving an incoming $b$ quark with momentum $p$ and outgoing $c$ quark
with momentum $p^\prime$, i.e. $\bar{u}_c(p^\prime)\Gamma u_b(p)$, where either
$\Gamma = \gamma_\mu$ or $\Gamma = \gamma_\mu \gamma_5$.
At one loop the graphs in Figures~\ref{ContVert} to \ref{ContLeg2} contribute.
The calculation was performed in Minkowski space using dimensional regularisation in Feynman gauge,
the on-shell $\overline{MS}$ scheme, and having introduced a fictitious gluon
mass $\lambda$ to regulate infra-red divergences. The treatment allowed
for non-zero external momenta,
 keeping terms linear in $\frac{\vec{p}}{M}$ and $\frac{\vec{p}^\prime}{M}$.

In general, we write the perturbative expansion of some quantity, $Q$, as 
\begin{equation}
Q = \sum\limits_{k=0}^\infty Q^{[k]} \alpha_s^k.
\end{equation}

We write the Feynman graph in Figure~\ref{ContVert} as $\alpha_s \Lambda^{\Gamma^{[1]}}$
and the leg corrections in Figures~\ref{ContLeg1} and \ref{ContLeg2} as
$\half \as \Gamma Z_{\psi_b}^{[1]}$ and $\half \as \Gamma Z_{\psi_c}^{[1]}$ respectively.
Dropping the external spinors, the total current correction may be written as

\begin{equation}
\Delta \Gamma = \as \left[ \Lambda^{\Gamma^{[1]}} + \half \Gamma  Z_{\psi_c}^{[1]} + \half Z_{\psi_b}^{[1]}\right],
\end{equation}
Here both $\Lambda^{\Gamma^{[1]}}$, and $Z_\psi$ contain (cancelling) infra-red divergences of modulus
$\frac{\as}{3\pi}4\log \frac{\lambda}{M}$. The result is IR-finite, and
we obtain that the $\overline{MS}$ current to one loop and $O(\frac{\vec{p}}{M})$ is as follows
\footnote{
It is worth noting here that at higher orders in the \pom expansion 
the coefficients $C_i$ become functions of the recoil 
$\omega=\frac{p\cdot p^\prime}{m_c m_b}$.}

\begin{equation}
\Delta \Gamma \cdot \Gtilde = C_1 \gamma^\mu + C_2 \frac{p^\mu}{m_b}  + C_3 \frac{p^{\prime\mu}}{m_c} 
\end{equation}
where,
\begin{eqnarray}
C_1 &=& \frac{\alpha_s}{3\pi} \left[ 
-3\frac{1+\xi}{1-\xi}\log\xi + 2 {\cal C} - 6
\right] \\
C_2 &=& \frac{\alpha_s}{3\pi} \left[
-\frac{4}{(1-\xi)^2}+2\xi\log\xi\frac{\xi-3}{(1-\xi)^3} + {\cal C} \left\{
2\frac{1+\xi}{(1-\xi)^2} + \frac{4\xi\log\xi}{(1-\xi)^3}
\right\}
\right] \\
C_3 &=& \frac{\alpha_s}{3\pi} \left[
2 \xi \frac{1+\xi}{(1-\xi)^2} + 4 \xi^2 \frac{\log\xi}{(1-\xi)^3} + {\cal C} \left\{ 
-\frac{4\xi^2}{(1-\xi)^2} + 2\xi \log\xi\frac{1-3\xi}{(1-\xi)^3}
\right\}
\right] 
\end{eqnarray}
and
\begin{equation}
\begin{array}{ccccc}
\xi = \frac{m_c}{m_b} &,&
{\cal C} = 
\left\{ 
\begin{array}{cl} 
 1  &; \Gamma = \gamma^\mu \\
-1  &; \Gamma = \gamma^\mu\gamma_5
\end{array} \right.
&,& 
\Gtilde = \left\{
\begin{array}{cl} 
 1        &; \Gamma = \gamma^\mu \\
 \gamma_5 &; \Gamma = \gamma^\mu\gamma_5
\end{array} \right.
\end{array}
\end{equation}

The Gordon decomposition of the spinor gives that, for $m_c=m_b$,
\begin{equation}
\bar{u} \gamma^\mu u = \frac{(p+p^\prime)^\mu}{2m} \bar{u} u
\end{equation}
and so the Vector current between degenerate quarks
is, of course, not renormalised.

In order to match these currents with Euclidean lattice NRQCD currents we 
first analytically continue this result to Euclidean space, then 
apply the Foldy-Wouthuysen expansion of the external spinors
\cite{Foldy} to the appropriate order, taking care with the Euclideanised
$\gamma$-matrices.

\subsection{Analytic Continuation}
We define the Euclideanisation as follows. Denoting the Minkowski 
gamma matrices and four vectors with a caret, $\hat{\gamma}$, so that
$\{\hat{\gamma}_\mu,\hat{\gamma}_\nu\} = 2 g_{\mu\nu}$, we use the Dirac-Pauli
representation:
\begin{equation}
\begin{array}{ccccc}
\hat{\gamma}^0 & = & \hat{\gamma}_0 & = & \left(\begin{array}{cc} I & 0\\ 0 & -I \end{array}\right) \\
\hat{\gamma}^j & = & -\hat{\gamma}_j &= & \left(\begin{array}{cc} 0 & \sigma_j\\ -\sigma_j & 0 \end{array}\right) 
\end{array}
\end{equation}
We also define the Euclidean four vectors in terms of the contravariant and covariant Minkowski
vectors as:
\begin{equation}
\begin{array}{ccccccc}
x^0 &=& x_0 &=& i \hat{x}^0 &=& i\hat{x}_0\\
x^j &=& x_j &=& \hat{x}^j &=& -\hat{x}_j
\end{array}
\end{equation}
We introduce Euclidean $\gamma$ matrices such that $\{\gamma_\mu,\gamma_\nu\} = 2 \delta_{\mu\nu}$
\begin{equation}
\begin{array}{ccccccc}
\gamma^0 &=& \gamma_0 &=& \hat{\gamma}^0 &=& \hat{\gamma}_0\\
\gamma^j &=& \gamma_j &=& - i \hat{\gamma}^j &=& i \hat{\gamma}_j\\
\gamma_5 &=& \hat{\gamma}_5
\end{array}
\end{equation}
For consistent notation in Euclidean space, we define Euclidean bilinear currents
in terms of the Euclidean $\gamma$ matrices, so that 
\begin{equation}
\begin{array}{ccccccc}
A^0 &=& A_0 &=&   \hat{A}^0 &=&  \hat{A}_0\\
A^j &=& A_j &=& -i\hat{A}^j &=& i\hat{A}_j\\
V^0 &=& V_0 &=&   \hat{V}^0 &=&  \hat{V}_0\\
V^j &=& V_j &=& -i\hat{V}^j &=& i\hat{V}_j
\end{array}
\end{equation}

Dropping terms in $A_k$ which are subsequently of higher order in
$\frac{\vec{p}}{M}$, and suppressing the spinors, this gives us in Euclidean space:
\begin{eqnarray}
\Delta V^0 &=& \gamma^0 \frac{\alpha_s}{3\pi} 
\left[ -3 \frac{1+\xi}{1-\xi}\log \xi - 4\right]  - 2 \frac{\alpha_s}{3\pi}\\
\Delta V^k &=& \gamma^k \frac{\alpha_s}{3\pi} \left[ -3 \frac{1+\xi}{1-\xi} \log \xi - 4 \right]
- i \frac{p^k}{m_b} \frac{\alpha_s}{3\pi}\left[ -\frac{2}{1-\xi} - \frac{2\xi \log \xi}{(1-\xi)^2} \right] 
- i \frac{p^{\prime k}}{m_c}\frac{\alpha_s}{3\pi} \left[ \frac{2\xi}{1-\xi} + \frac{2\xi \log \xi}{(1-\xi)^2} \right] 
\\
\Delta A^0 &=& \gamma^0 \gamma_5 \frac{\alpha_s}{3\pi} \left[ -3\frac{1+\xi}{1-\xi}\log\xi - 8 \right]
+ \gamma_5 \frac{\alpha_s}{3\pi} \left[ - 6 \frac{1+\xi}{1-\xi} - 12 \frac{\xi \log \xi}{(1-\xi)^2} \right]
\\
\Delta A^k &=& \gamma^k\gamma_5 \frac{\alpha_s}{3\pi} \left[ -3\frac{1+\xi}{1-\xi}\log\xi - 8 \right] .
\end{eqnarray}

\subsection{Non-Relativistic Expansion}

We now perform the Foldy-Wouthuysen expansion of the external spinors $u_h$ to
order $\frac{1}{m}$ in terms of the Euclidean $\gamma$ matrices to produce
two component spinors $U_Q$:
\begin{equation}
u_h(p) = \left[ 1 - \frac{i}{2m} \vec{\gamma} \cdot \vec{p} \right] u_Q
\end{equation}
where
\begin{equation}
u_Q = \left[\begin{array}{c} U_Q \\ 0 \end{array} \right]
\end{equation}
and $U_Q$ is a two component spinor, and $m$ is the pole mass of the quark arising from
the on-shell condition (more formally, $m$ is the one loop pole mass whenever it appears 
at tree level since we are working consistently to order one loop). 
Inserting this in the results for the continuum renormalisation and keeping terms to order 
$\frac{1}{M}$ results in the following mixed continuum currents, with the relevant two component
spinors suppressed.
\begin{eqnarray}
\label{ContV0ren}
V^0 &=& 1_{\rm pauli} \left[ 1 + \frac{\alpha_s}{3\pi} \left( -3 \frac{1+\xi}{1-\xi}\log \xi - 6 \right) \right] \\
V^k &=& 
\left[ -i \sigma_k \frac{\sigma\cdot p}{2m_b}  - \frac{\sigma \cdot p^\prime}{2m_c} i \sigma_k \right]
\left[ 1 + \frac{\alpha_s}{3\pi} \left( -3 \frac{1+\xi}{1-\xi}\log \xi  - 4 \right) \right] 
\\
\nonumber && -i \frac{p_k}{m_b} \frac{\alpha_s}{3\pi} 
\left[ -\frac{2}{1-\xi} -\frac{ 2 \xi \log \xi}{(1-\xi)^2}\right] 
\\
\nonumber && -i \frac{p_k^\prime}{m_c} \frac{\alpha_s}{3\pi} 
\left[ \frac{ 2 \xi}{1-\xi} + \frac{ 2 \xi \log \xi}{(1-\xi)^2} \right] \\
A^0 & = & 
\left[ \frac{\sigma\cdot p}{2m_b} + \frac{\sigma\cdot p^\prime}{2m_c} \right] 
\left[ 1 + \frac{\alpha_s}{3\pi} \left( -3\frac{1+\xi}{1-\xi}\log\xi - 8 \right) \right] \\
\nonumber&&+ 
\left[ \frac{\sigma\cdot p}{2m_b} - \frac{\sigma\cdot p^\prime}{2m_c} \right] 
\frac{\alpha_s}{3\pi} \left[  - 6 \frac{1+\xi}{1-\xi} - 12 \frac{\xi \log \xi}{(1-\xi)^2} \right]\\
\label{ContAkren}
A_k &=& -i \sigma_k 
\left[ 1 + \frac{\alpha_s}{3\pi} \left( -3 \frac{1+\xi}{1-\xi} \log \xi - 8 \right)\right] 
\end{eqnarray}

We therefore define continuum Euclidean operators $\Omega_i^J$ in Table~\ref{TabContOps},
to which we wish to match a corresponding set of lattice operators,
and use the notation 
\begin{equation}
\langle c(p^\prime)| J^{\overline{MS}}| b(p) \rangle^{\rm 1-loop} = 
\sum\limits_\alpha Z^{\overline{MS}}_\alpha U_c^\dagger \Omega_\alpha^J U_b .
\end{equation}

The vector current is a conserved current for $m_b = m_c$ and, for
this reason, the continuum renormalisation ( factor in square brackets in (\ref{ContV0ren}) ) is unity in
the $\xi\rightarrow 1$ limit.
The operator corresponding to the temporal component of the vector current, $1_{\rm pauli}$,
is a conserved current of both the lattice NRQCD and HQET actions, and we shall 
see that the corresponding lattice renormalisation is also trivial in the $m_b = m_c$ limit. 

\label{ContBLM}
Unlike the lattice correction, 
the pure HQET contribution to the vector current renormalisation at zero recoil remains trivial
with non-identical quark masses at leading order in $\frac{1}{m_b}$ and $\frac{1}{m_c}$. As a result 
the above expressions (\ref{ContV0ren}) and (\ref{ContAkren})
are the same as the renormalisation at zero recoil 
for the continuum heavy quark effective theory, 
commonly referred to as $\eta_V$ and $\eta_A$ respectively
\cite{Neubert}. This is somewhat fortuitous since it will allow us, in section \ref{SecScale},
to use the existing result 
for the BLM scale $\mu$ \cite{BLMscale,BLMEtaVscale} for the processes contributing to expressions 
(\ref{ContV0ren}) and (\ref{ContAkren}),
with $\mu_V \simeq 0.92 \sqrt{m_b m_c}$ and $\mu_A \simeq 0.51 \sqrt{m_b m_c}$.

\section{Lattice Calculation}

\label{SecLattice}
\subsection{Lattice action, currents, and Feynman rules}

We use the usual Wilson plaquette action \cite{WilsonGauge} defined
in terms of the usual SU(3) gauge links $U_\mu(x)$,
and the tadpole improved \cite{LepageMackenzie} $O(\frac{1}{M})$ lattice NRQCD action \cite{LepageNRQCD},
\begin{equation}
\begin{array}{c}
a {\cal L}_{\rm NRQCD} = \psi^\dagger(x) \psi(x) 
- \psi^\dagger(x+a \hat{t} ) \left( 1-\frac{a\delta H}{2}\right)
\left( 1 - \frac{a H_0}{2n}\right)^n \frac{U_t^\dagger(x)}{u_0} 
\left( 1 - \frac{a H_0}{2n}\right)^n\left( 1-\frac{a\delta H}{2}\right) \psi(x)
\end{array}
\end{equation}
where
\begin{eqnarray}
H_0 &=& -\frac{\Delta^2}{2M_0},\\
\delta H &=& -c_B  \frac{g}{2M_0} \sigma\cdot B,
\end{eqnarray}
\begin{equation}
\Delta^2 \chi(x) = \frac{1}{u_0} \sum\limits_{i=x,y,z} \left( 
U_i(x) \chi(x+\hat{i}) + U^\dagger_i(x-\hat{i}) \chi(x-\hat{i}) - 2 u_0 \chi(x)
\right),
\end{equation}
and
\begin{eqnarray}
B_i(x) &=& -\frac{1}{2} \epsilon_{ijk} \left\{ {\cal F}_{jk}(x) - \frac{1}{3} {\rm Tr} {\cal F}_{jk}(x)\right\},\\
{\cal F}_{\mu\nu}(x) &=& \frac{1}{4 ag^2} \sum\limits_{\alpha=\pm \mu, \beta=\pm \nu}
 \frac{\left[ P_{\alpha\beta}(x) - P_{\alpha\beta}^\dagger(x)\right]}{2i},\\
P_{\alpha\beta}(x) &=& \frac{1}{u_0^4} U_\alpha(x)U_\beta(x+\hat{\alpha})U^\dagger_\alpha(x+\hat{\beta})U^\dagger_\beta(x).
\end{eqnarray}

In the final results quoted in this paper, $u_0$ is taken from the plaquette and 
to one loop order in perturbation theory $u_0^{[1]} = - \frac{\pi}{3}$. However, sufficient
information will presented to allow a reader to change these results for different
values of $u_0$. 
To this order in $\as$ any dynamical quarks present in the simulation
do not appear in the perturbation theory. This calculation is therefore applicable to both
quenched and dynamical simulations. It is well known how to perform
perturbative calculations of the renormalisation of weak currents in NRQCD \cite{JunkoColin},
and the Feynman rules in the Euclidean lattice theory used here are the
same as those published by Morningstar and Shigemitsu \cite{JunkoColin}.
This paper completes a programme of work previously discussed at 
two lattice conferences \cite{PBCTHD}.

Inspection of the continuum currents suggests the basis for the
lattice current operators given in Table~\ref{TabLatOps}. 
The momentum dependent operators
are reproduced on the lattice with gauge covariant derivatives,
\begin{equation}
\nabla_\mu \chi(x) = \frac{1}{2 u_0}\left[ 
U_\mu(x) \chi(x+\hat{\mu}) - U_\mu^\dagger(x-\hat\mu)\chi(x-\hat\mu)\right]
\end{equation}
which introduces additional quark gluon vertices that are not
present in the continuum theory, as well as a tadpole improvement counter term.
For convenience in simulations, we define the
lattice operators in terms of the bare mass, $M_q$. Since we require that the one-loop 
renormalised kinetic mass in NRQCD matches the one-loop renormalised mass in the 
relativistic theory 
\footnote{Lorentz
invariance implies that the kinetic and pole masses are consistent in the relativistic theory}, 
we must introduce additional $Z_m$ factors for lattice currents wherever the bare mass 
$M_q$ appears at tree level.
The mass ratio $\xi$ appears in the continuum coefficients only at
order $\as$, and hence the bare ratio $\frac{M_c}{M_b}$ can be used to this order.

\begin{figure}
\caption{
\label{FigCurrentInsertions} 
Feynman rules associated with insertion of a lattice weak current 
operator to order $g^0$, $g^1$, and $g^2$. The gluon couplings
only arise for $\Gamma = V_i$ and $\Gamma = A_0$. Here $\beta$ runs over
the dimension of the lattice operator basis corresponding to the continuum
current $\Gamma$, as given in Table~\ref{TabLatOps}. 
}
\setlength\epsfxsize{200pt} \setlength\epsfysize{150pt} \epsfbox{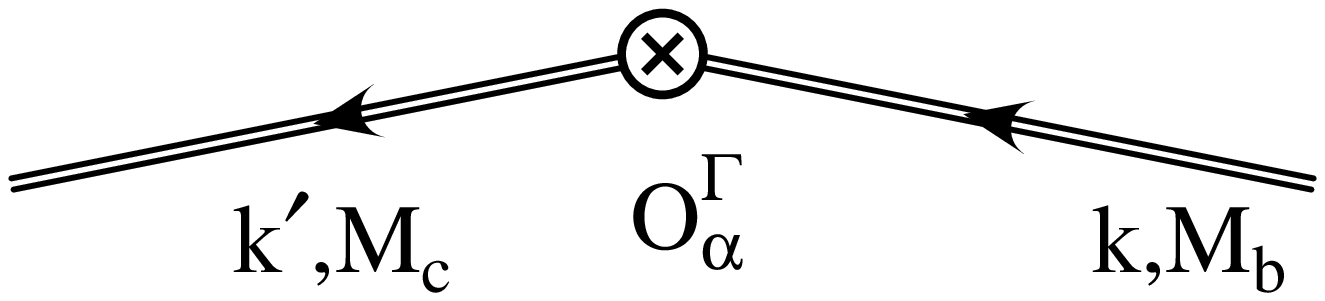} 
\setlength\epsfxsize{200pt} \setlength\epsfysize{150pt} \epsfbox{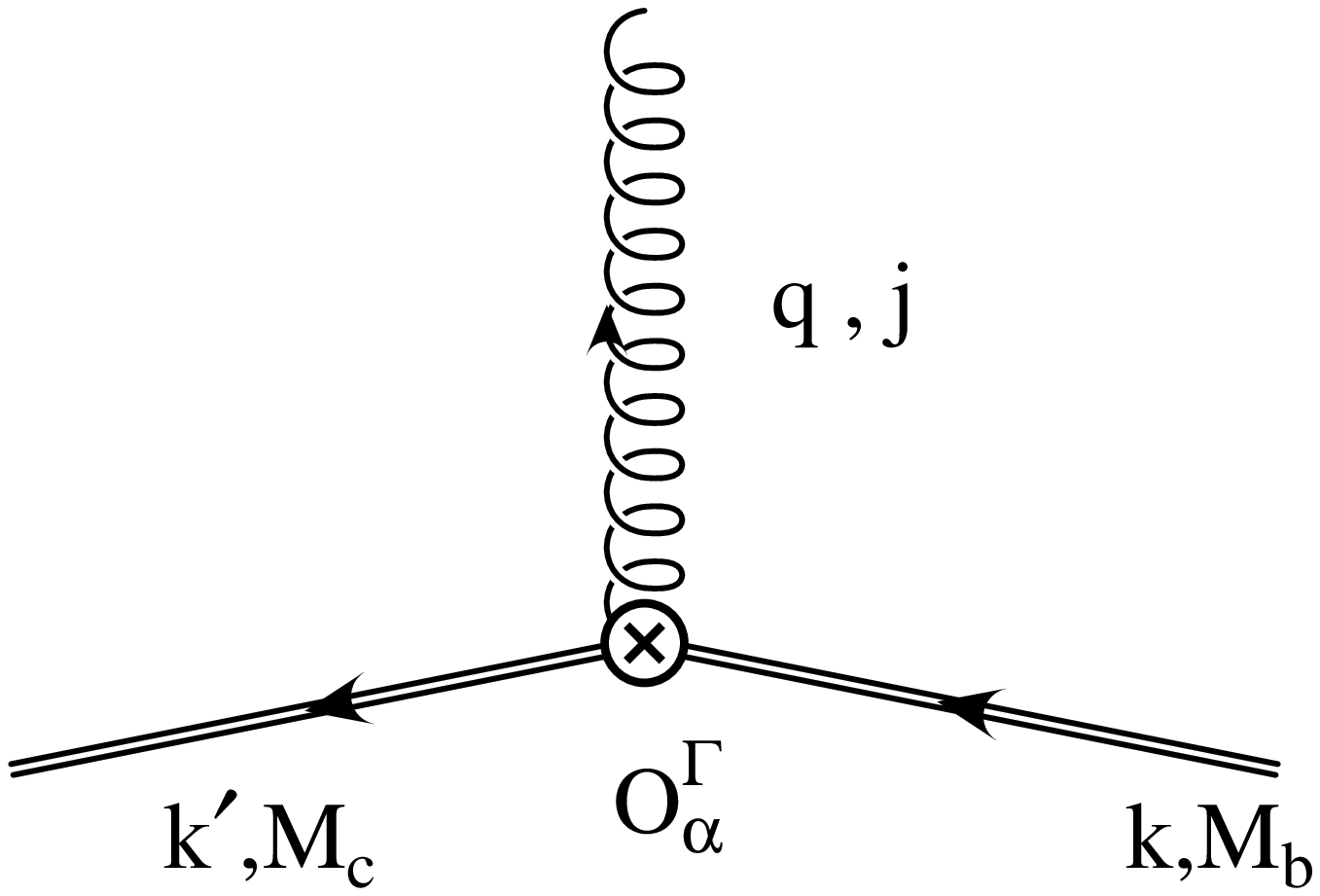} 
\setlength\epsfxsize{200pt} \setlength\epsfysize{150pt} \epsfbox{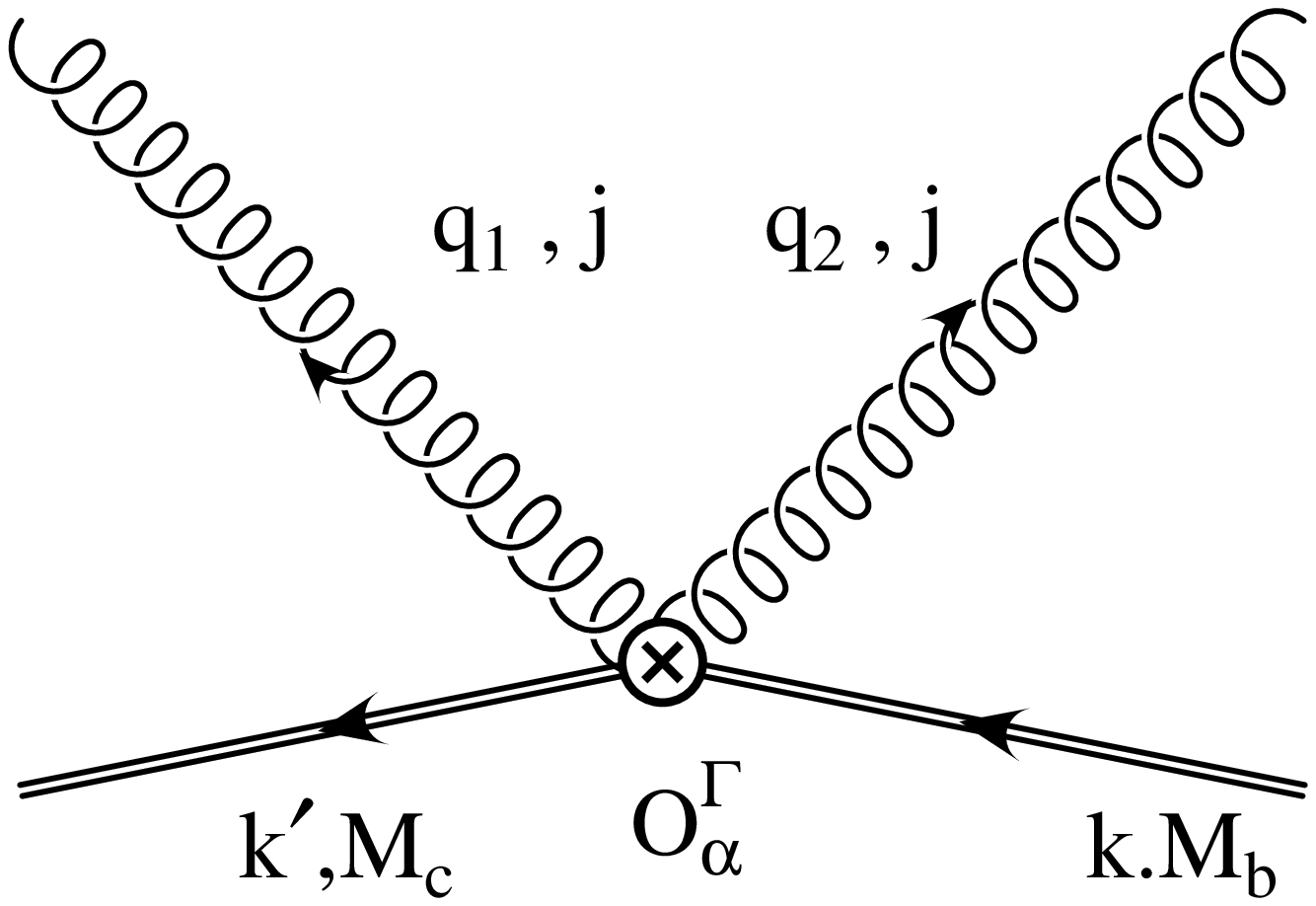}
\end{figure}

We have additional Feynman rules associated with the 
current operators involving derivatives (and hence gauge links), 
giving both one and two gluon couplings. 
The conventions for momenta involved in current insertions of an operator ${\cal O}$ 
both at leading order, and couplings to both 1 and 2 gluons of polarisation $j$
are as given in figure \ref{FigCurrentInsertions}. 

The Feynman rules associated with the spatial vector and temporal axial
currents at the vertex are simply the insertion of the appropriate Pauli structure.
The Feynman rules associated with the spatial vector current insertions 
are as follows
\begin{eqnarray}
{\cal O}_1^{V_k} ({\rm tree})   &=& -\frac{i}{2M_b}\sum\limits_j \sin k_j \sigma_k\sigma_j\\
{\cal O}_1^{V_k} ({\rm 1-gluon}) &=& -\frac{i g}{2M_b} \cos \left[ k_j - \half q_j\right]\sigma_k\sigma_j\\ 
{\cal O}_1^{V_k} ({\rm 2-gluon}) &=& \frac{i g^2}{4M_b} \sin \left[ k_j - \half (q_{1_j}+q_{2_j}) \right]\sigma_k\sigma_j
\end{eqnarray}

\begin{eqnarray}
{\cal O}_3^{V_k} ({\rm tree})    &=& -\frac{i}{M_b}\sin k_k \\
{\cal O}_3^{V_k} ({\rm 1-gluon}) &=& -\frac{i g}{M_b}\cos \left[ k_j - \half q_j\right]\delta_{jk} \\ 
{\cal O}_3^{V_k} ({\rm 2-gluon}) &=& \frac{i g^2}{2 M_b}\sin \left[ k_j - \half (q_{1_j}+q_{2_j}) \right]\delta_{jk} 
\end{eqnarray}
where the 1 and 2 gluon couplings are to gluons of polarisation $j$.

The rules associated with the leftward derivative currents are both the obvious
analogues of the above and not in fact required since the diagrams involving the
second current may be switched for those of the first current with the masses
swapped and momenta reversed, halving the required coding effort. 
The Feynman rules associated with the first temporal axial lattice current
are as follows,
\begin{eqnarray}
{\cal O}_1^{A_0} ({\rm tree})    &=& \frac{1}{2M_b} \sum\limits_j \sin k_j \sigma_j \\
{\cal O}_1^{A_0} ({\rm 1-gluon}) &=& \frac{g}{2M_b} \cos \left( k_j - \half q_j \right) \sigma_j\\
{\cal O}_1^{A_0} ({\rm 2-gluon}) &=& \frac{g^2}{4M_b} \sin \left( k_j - \half (q_{1j} + q_{2j}) \right) \sigma_j
\end{eqnarray}
while, as above, the rule for the second current is not required.

\subsection{Mixing Matrix}

\begin{figure}[htb]
\caption{Regular topology vertex correction.\\
Temporally polarised gluons couple through the $\partial_t$ term in the action for 
and spatially polarised gluons through both the $\frac{p^2}{2m}$ 
and $\frac{\sigma \cdot B}{2m}$ term. Graphs involving mixed $\frac{p^2}{2m}$ and
$\frac{\sigma \cdot B}{2m}$ couplings vanish for the temporal vector and spatial
axial currents.
}
\label{FIGvertex_reg}
\setlength\epsfxsize{150pt}
\epsfbox{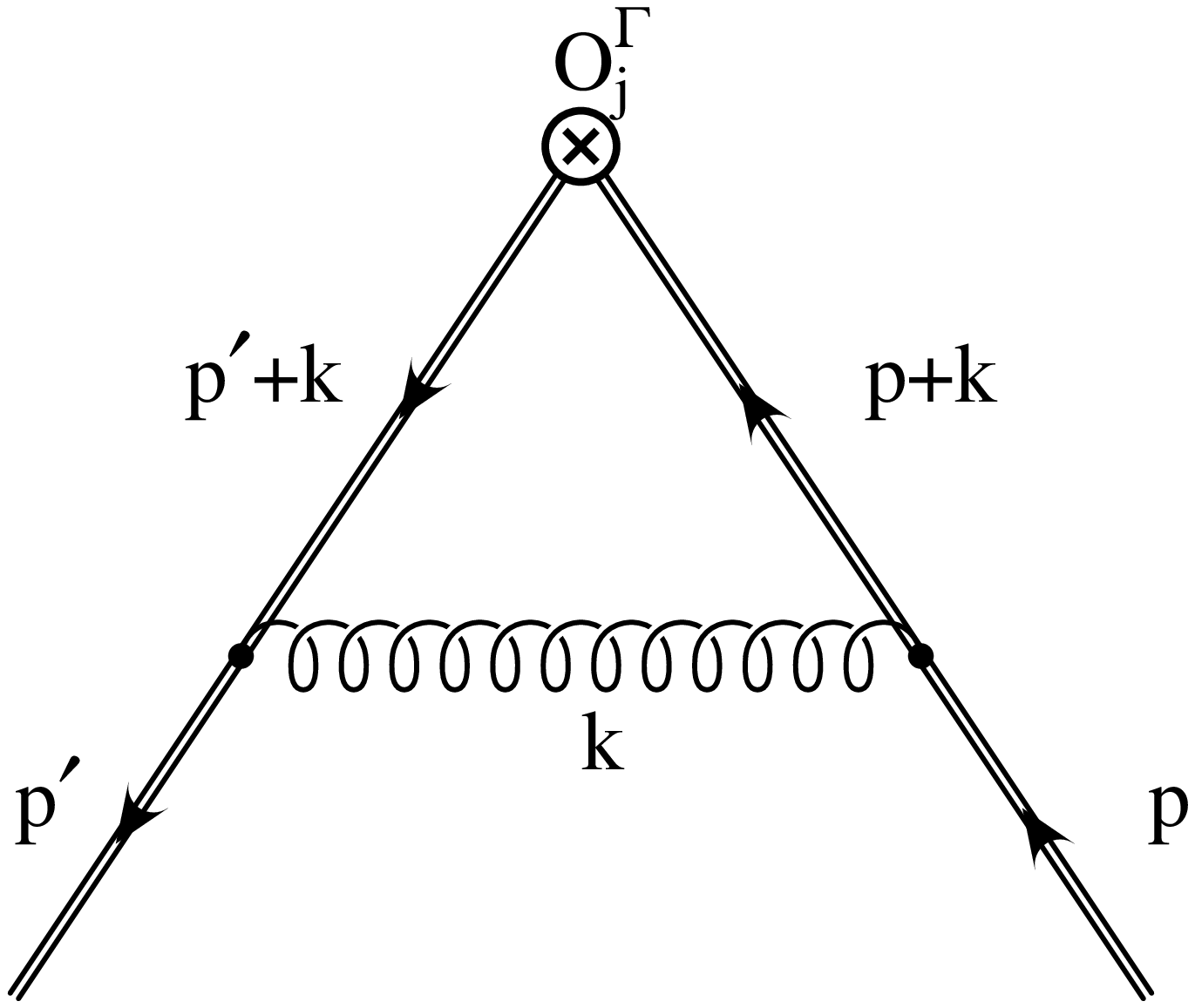}
\end{figure}

We define the 1-loop mixing matrix of the lattice theory for each of
the four currents $J\in\{V_0,V_k,A_0,A_k\}$ as follows:
\begin{equation}
\langle c(p^\prime)| {\cal O}^J_\alpha | b(p) \rangle^{\rm 1-loop}
 = \sum\limits_\beta {\cal M}^J_{\alpha\beta} U_c^\dagger \Omega_\beta^J U_b ,
\end{equation}
where the range of the summation over $\beta$ depends on the dimension of the 
basis for each continuum current.
To avoid ambiguity, we shall denote the symbols for lattice Feynman diagrams in bold font to
differentiate them from symbols used in the continuum calculation
(although the different symbols do not in fact appear in the same equations).
For both the temporal vector and spatial axial currents the mixing matrix
is rank one, and we have (where $J$ is either $V_0$ or $A_k$):

\begin{equation}
{\cal M}^{J} = 1 + \alpha_s \left[  
\half\{ {\bf Z}_{\psi_c}\oneloop + {\bf Z}_{\psi_b}\oneloop \}
+ {\bf \Lambda}^{J \soneloop}
\right] ,
\end{equation}
${\bf Z}_{\psi}\oneloop$ gives the relevant leg correction, whose relation
to the self energy diagram (Figure~\ref{FIGself_reg}) is described in appendix~\ref{AppA},
where we have followed closely the conventions of Morningstar~\cite{ColinSelfEnergy}.
${\bf \Lambda}^{J \soneloop}$ corresponds to contributions from
the Feynman diagram of Figure~\ref{FIGvertex_reg} where gluon couplings
from each of the $\nabla_t$, $\frac{p^2}{2m}$, and $\frac{\sigma\cdot {\bf B}}{2m}$
terms in the action are included at each vertex.

\begin{figure}[htb]
\caption{Diagrams of ''left ear'' and ``right ear''topologies contribute to both the temporal axial and
spatial vector currents. The gauge link
in the spatial derivative in the current operators creates a coupling to spatially
polarised gluons, and the coupling to the leg arises from both the $\frac{p^2}{2m}$
and $\frac{\sigma \cdot B}{2m}$ terms. 
}
\label{FIGvertex_left_ear}
\setlength\epsfxsize{150pt}
\epsfbox{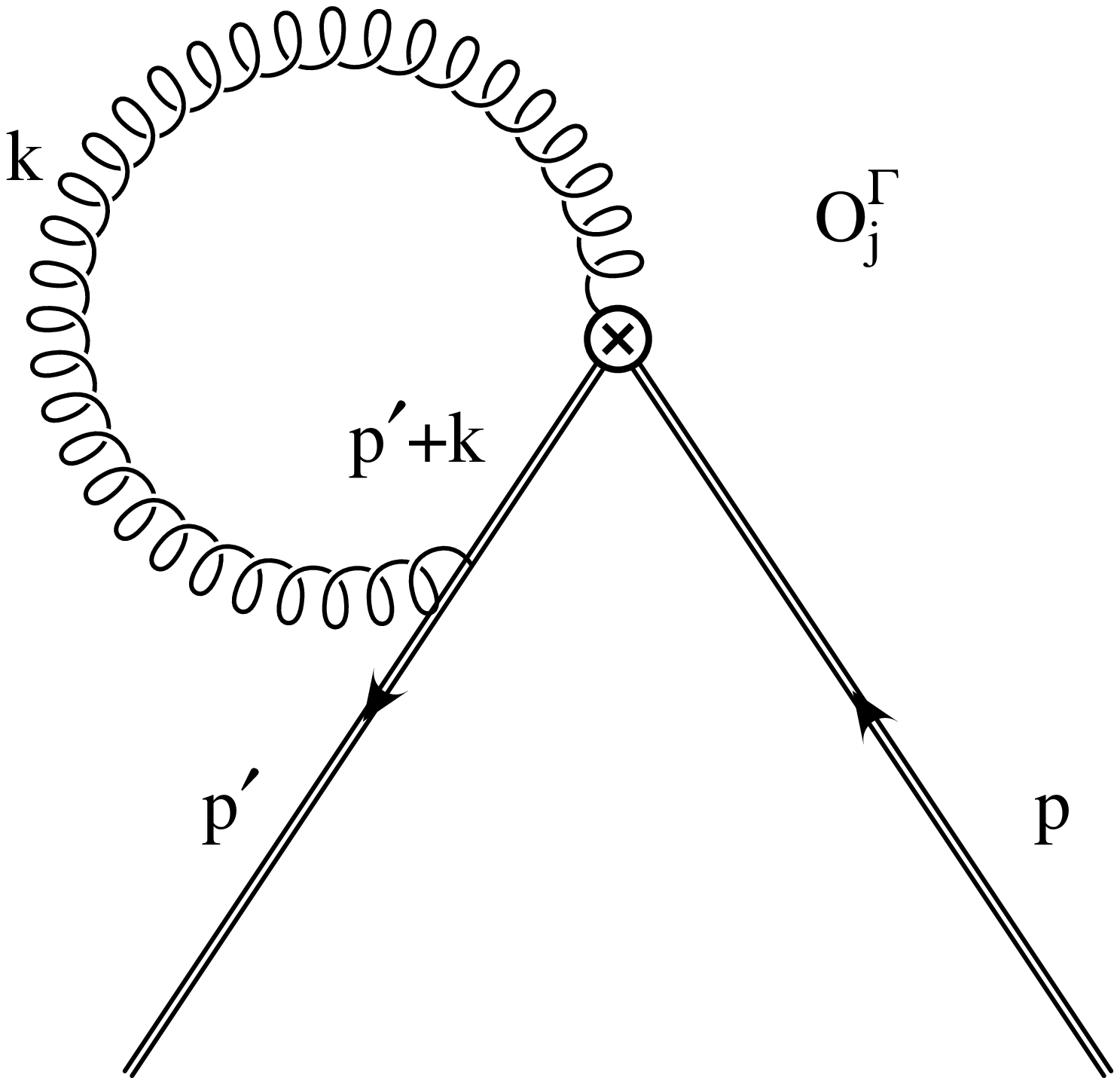}
\label{FIGvertex_right_ear}
\setlength\epsfxsize{150pt}
\epsfbox{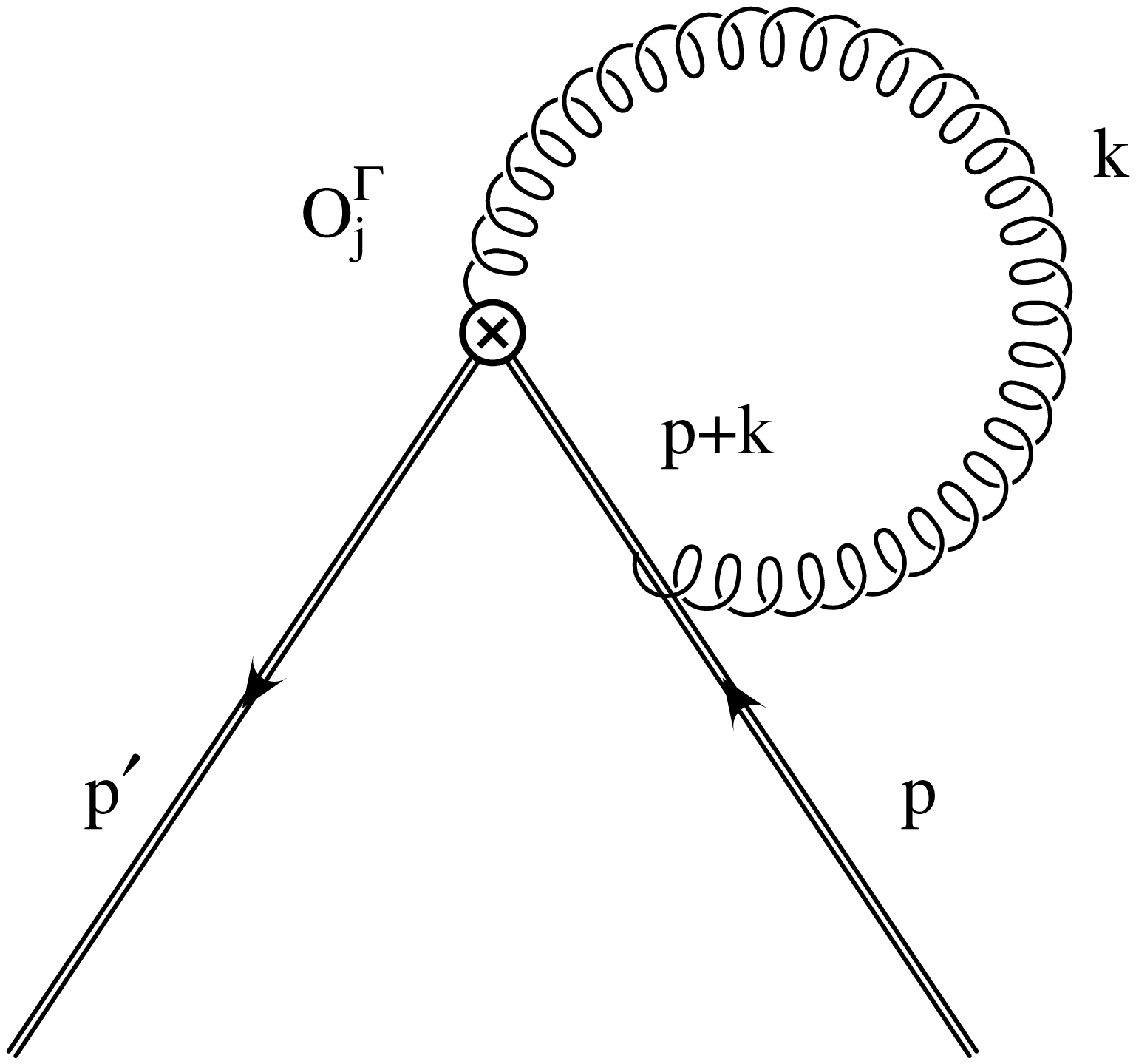}
\end{figure}

For the spatial vector current (and indeed the temporal axial current) 
the mixing matrix is non-trivial. We denote the vertex correction to each of the lattice
currents ${\cal O}^{V_k}_\alpha$ from the sum of Feynman diagrams represented in  
Figures~\ref{FIGvertex_reg} and~\ref{FIGvertex_right_ear} by ${\bf \Lambda}^{V_k\soneloop}_\alpha$.
At one loop the the vertex correction overlaps
with each of the corresponding continuum operators, giving rise 
to the non-trivial mixing problem
\begin{equation}
{\bf \Lambda}^{V_k\soneloop}_\beta = \sum\limits_\beta {\bf \Lambda}^{V_k\soneloop}_{\alpha\beta} \Omega_\beta^{V_k}.
\end{equation}
We resolve the Pauli and momentum structure of ${\bf \Lambda}^{V_k\soneloop}_1$ using the following
projections with $j \ne k$, then setting the spatial momenta to zero:
\begin{equation}
\begin{array}{cc}
{\bf \Lambda}^{V_k \soneloop}_{11} = 2 i M_b \sigma_j \sigma_k \frac{\partial}{\partial p_j} 
{\bf \Lambda}_1^{V_k \soneloop} &
{\bf \Lambda}^{V_k \soneloop}_{12}= 2 i M_c \sigma_k \sigma_j \frac{\partial}{\partial p^\prime_j} 
{\bf \Lambda}_1^{V_k \soneloop} \\
{\bf \Lambda}^{V_k \soneloop}_{13}=
i M_b \frac{\partial}{\partial p_k}{\bf \Lambda}_1^{V_k \soneloop} -\half {\bf \Lambda}_{11}^{V_k \soneloop} &
{\bf \Lambda}^{V_k \soneloop}_{14}=
i M_c \frac{\partial}{\partial p^\prime_k}{\bf \Lambda}_1^{V_k \soneloop}-\half {\bf \Lambda}_{12}^{V_k \soneloop}.
\end{array}
\end{equation}
We may use the bare lattice mass for convenience in place of the
the pole mass in these projections 
since we are evaluating one-loop coefficients, the difference
being $O(\as^2)$.
Mixing matrix elements ${\bf \Lambda}_{2\beta}$ are related to ${\bf \Lambda}_{1\beta}$ by 
$M_b \leftrightarrow M_c$, and it will be seen later that ${\bf \Lambda}_{3\beta}$ and ${\bf \Lambda}_{4\beta}$ 
are not required for our calculation. Hence we write
the contributions to the mixing matrix for the spatial vector current as follows,
\begin{equation}
{\cal M}^{V_k}_{\alpha\beta} = \delta_{\alpha\beta} + \alpha_s
\left[ \begin{array}{c}
\half\{ {\bf Z}_{\psi_c}\oneloop + {\bf Z}_{\psi_b}\oneloop\} \delta_{\alpha\beta}
+ \left\{{\bf Z}_{m_b}\oneloop+{\bf Z}_{m_b}^{\rm TI \soneloop}\right\} \left\{\delta_{\alpha 1}+\delta_{\alpha 3}\right\} \delta_{\alpha\beta}\\
+ \left\{{\bf Z}_{m_c}\oneloop+{\bf Z}_{m_c}^{\rm TI \soneloop}\right\} \left\{\delta_{\alpha 2}+\delta_{\alpha 4}\right\} \delta_{\alpha\beta}
+ {\bf \Lambda}_{\alpha\beta}^{V_k \soneloop} + {\bf \Lambda}_{\alpha\beta}^{{\rm TI}^{V_k} \soneloop}
\end{array}
\right].
\end{equation}
where the the ${\bf Z}_m$ factors arise from matching the bare mass in the lattice current
to the pole mass in the continuum current to order $\alpha_s$, and the tadpole improvement
counter terms are
\begin{eqnarray}
{\bf Z}_m^{{\rm TI} \soneloop}          &=&  u_0^{[1]}\left( 1 - \frac{3}{2n aM_0} \right)\\
{\bf \Lambda}_{\alpha\beta}^{{\rm TI}^{V_k} \soneloop} &=& - \delta_{\alpha\beta} u_0^{[1]} .
\end{eqnarray}
Similarly, for the temporal axial current we project the vertex corrections
on to the continuum basis using,
\begin{equation}
\begin{array}{cc}
{\bf \Lambda}^{A_0 \soneloop}_{11}  = 2 M_b \sigma_j \frac{\partial}{\partial p_j} 
{\bf \Lambda}_1^{A_0 \soneloop}, &
{\bf \Lambda}^{A_0 \soneloop}_{12}  = 2 M_c \sigma_j \frac{\partial}{\partial p^\prime_j} 
{\bf \Lambda}_1^{A_0 \soneloop}.
\end{array}
\end{equation}
There is also a tadpole improvement counter term for the temporal axial current,
\begin{equation}
{\bf \Lambda}_{\alpha\beta}^{{\rm TI}^{A_0} \soneloop} = - \delta_{\alpha\beta} u_0^{[1]} .
\end{equation}
We write the contributions to the mixing matrix for the temporal axial current as follows,
\begin{equation}
{\cal M}^{A_0}_{\alpha\beta} = \delta_{\alpha\beta} + \alpha_s
\left[ \begin{array}{c}
\half\{ {\bf Z}_{\psi_c}\oneloop + {\bf Z}_{\psi_b}\oneloop\} \delta_{\alpha\beta}
+ \{{\bf Z}_{m_b}\oneloop + {\bf Z}_{m_b}^{{\rm TI} \soneloop} \} \delta_{\alpha 1} \delta_{\alpha\beta}\\
+ \{{\bf Z}_{m_c}\oneloop + {\bf Z}_{m_c}^{{\rm TI} \soneloop} \} \delta_{\alpha 2} \delta_{\alpha\beta}
+ {\bf \Lambda}_{\alpha\beta}^{A_0 \soneloop}+ {\bf \Lambda}_{\alpha\beta}^{{\rm TI}^{A_0} \soneloop}
\end{array}
\right].
\end{equation}
The major task in the renormalisation is to calculate the terms in the
above mixing matrices numerically using the Feynman rules for the lattice
action. 

The integrands for each of the Feynman graphs 
(and their first and second derivatives with respect to both
$p_i$ and $p^\prime_j$ where appropriate) 
were coded in two independent ways in order to verify the
correctness of the calculation. 

A C++ class was written to encode the Pauli algebra, 
and wrapped in a Taylor series class giving 
the first few entries of the (double) expansion in $p_j$ and $p^\prime_k$.
The standard operators were overloaded and standard functions applied
the chain rule to their arguments. This enabled the Feynman diagrams
to be constructed easily from the Feynman rules, with the Pauli manipulations
taken care of by the code. The leading entries in the double Taylor series
gave the first few derivatives with respect to $p_j$ and $p^\prime_k$.
Additionally much faster Fortran codes were written with the Pauli manipulations carried
out manually, and all derivatives with respect to the external momenta taken
by hand. VEGAS \cite{GPLvegas} was used to evaluate the one loop integrals.

The lattice effective theory is constructed to have the same infra-red physics as the 
relativistic continuum theory, so that
we find the same (cancelling) logarithmic infra-red divergences in the lattice
renormalisation in the lattice diagrams as were present in continuum theory.
These arise from graphs where a temporally polarised gluon is exchanged
via the $\nabla_t$ coupling in the action.
To handle the infra-red divergent graphs we
always perform the integral over $k_0$ analytically, and
then integrate infra-red finite sums of graphs, e.g.
$
\half\{{\bf Z}_{\psi_c}\oneloop+{\bf Z}_{\psi_b}\oneloop\} + {\bf \Lambda}_{11}\oneloop 
$,
over the remaining three dimensions corresponding to the spatial components of
the loop momentum.

\section{Matching and Results}

\label{SecMatching}

We wish to form a linear combination of the lattice currents 
such that they match the continuum currents at one loop.
We reconstruct the $\overline{MS}$ continuum currents 
from the lattice currents as follows, 

\begin{equation}
\begin{array}{ccc}
\langle c(p^\prime)|J^{\overline{MS}}| b(p) \rangle^{\rm 1-loop}
 &=& \sum\limits_{\alpha,\beta} Z^{\overline{MS}}_\alpha ({\cal M}^{J})^{-1}_{\alpha\beta} 
\langle c(p^\prime)|{\cal O}^J_\beta | b(p) \rangle^{\rm 1-loop}\\
&=& \sum\limits_\beta \left[ Z^{\rm tree}_\beta+\alpha_s\rho^J_\beta \right]\langle c(p^\prime)|{\cal O}^J_\beta | b(p) \rangle^{\rm 1-loop}
\end{array}
\end{equation}
where $Z^{\rm tree}_\beta\in\{0,1\}$ is the tree level piece of 
$Z^{\overline{MS}}_\beta$

In what follows we shall present results for the $\rho_\beta$ for each of the currents in turn
as a function of both the bare lattice
mass $\frac{1}{M_b}$ and the mass ratio $\xi = \frac{M_c}{M_b}$.
For reasons of practicality, we make the stabilisation parameters
for each quark, $n_c$ and $n_b$, implicit functions of the mass as
indicated in table~\ref{TABnhams}.

\subsection{Temporal Vector Current}
For the temporal vector current we have
\begin{eqnarray}
 V_0^{\overline{MS}}  &=&
\left\{ 1 + \alpha_s \rho^{V_0}\right\} {\cal O}^{V_0},\\
\rho^{V_0} &=&
B^{V_0} - \half\{ {\bf Z}_{\psi_c}\oneloop + {\bf Z}_{\psi_b}\oneloop \}
- {\bf \Lambda}^{V_0 \soneloop},\\
B^{V_0}&=& \frac{1}{3\pi}\left[ - 3\frac{1+\xi}{1-\xi}\log\xi - 6\right],
\end{eqnarray}

\begin{figure}
\caption{$\rho^{V_0}$ as a function of $\xi$ and $\frac{1}{M_b}$.
\label{FIGrhoV03d}}
\setlength\epsfxsize{350pt} \setlength\epsfysize{200pt}
\setlength\epsfxsize{250pt} \setlength\epsfysize{250pt}
\epsfbox{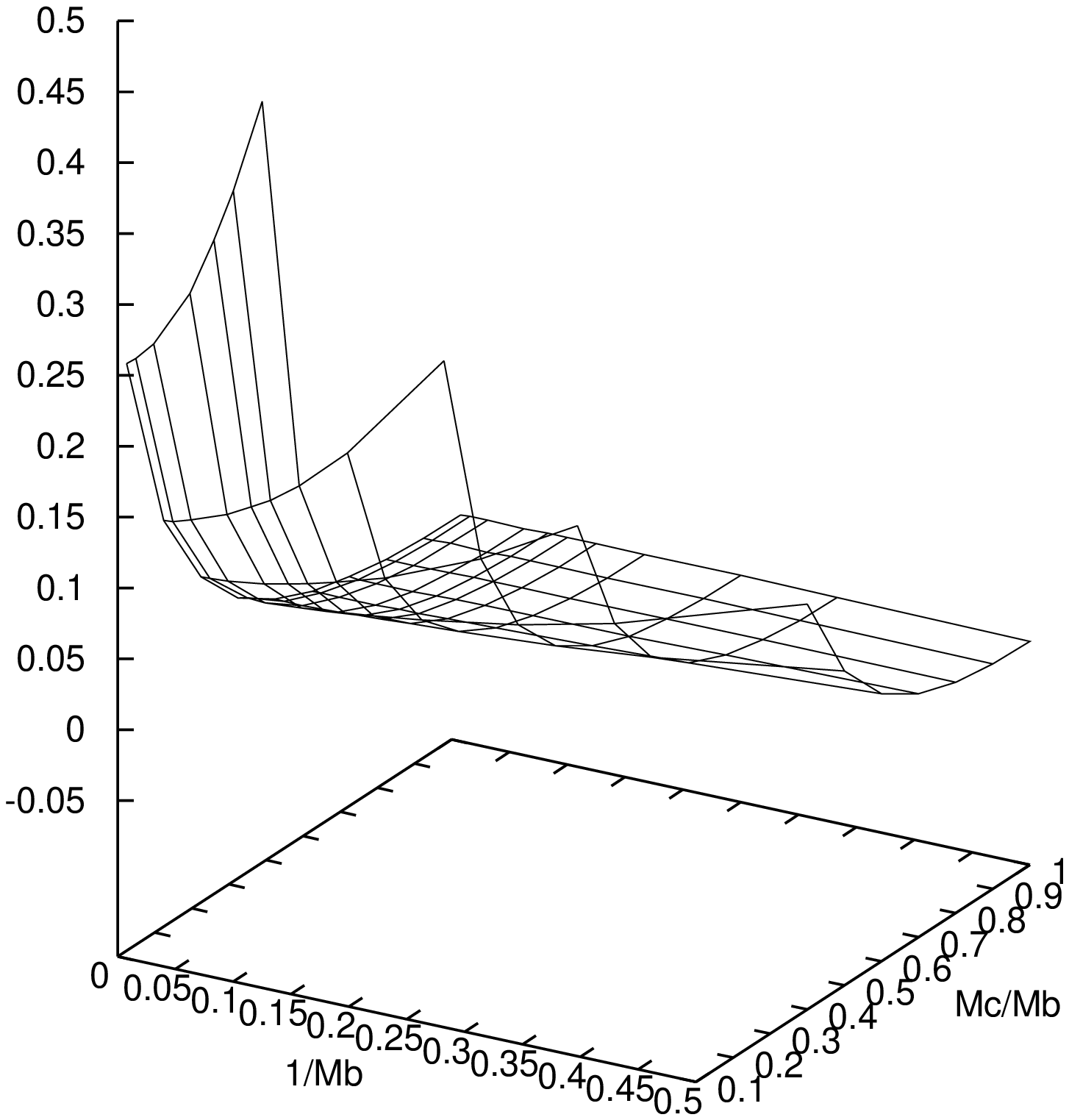}
\end{figure}

We plot the renormalisation coefficient $\rho^{V_0}$ in figure~\ref{FIGrhoV03d}.
In table~\ref{TABrhoV0} we give the correction as a function
of $\frac{1}{M_b}$ for $\xi = 0.1,0.2,0.3,0.4,0.5$ and $ 1.0$, 
useful ratios for lattice simulations. Since the lattice current is conserved, 
the one-loop renormalisation coefficient vanishes, for all $M_b$ in the case
$\xi = 1$, as clearly shown by figure~\ref{FIGrhoV03d}. 

\begin{figure}
\caption{Pure lattice correction $\rho^{V_0}-B^{V_0}$ as a function of $(\frac{1}{M_c}-\frac{1}{M_b})^2$ for curves $\xi = 0.2\ldots0.5$ running from top to bottom. The leading correction is linear and universal showing that Luke's
theorem is held to one-loop in lattice NRQCD.
\label{FIGluke}}
\setlength\epsfxsize{250pt} \setlength\epsfysize{250pt}
\epsfbox{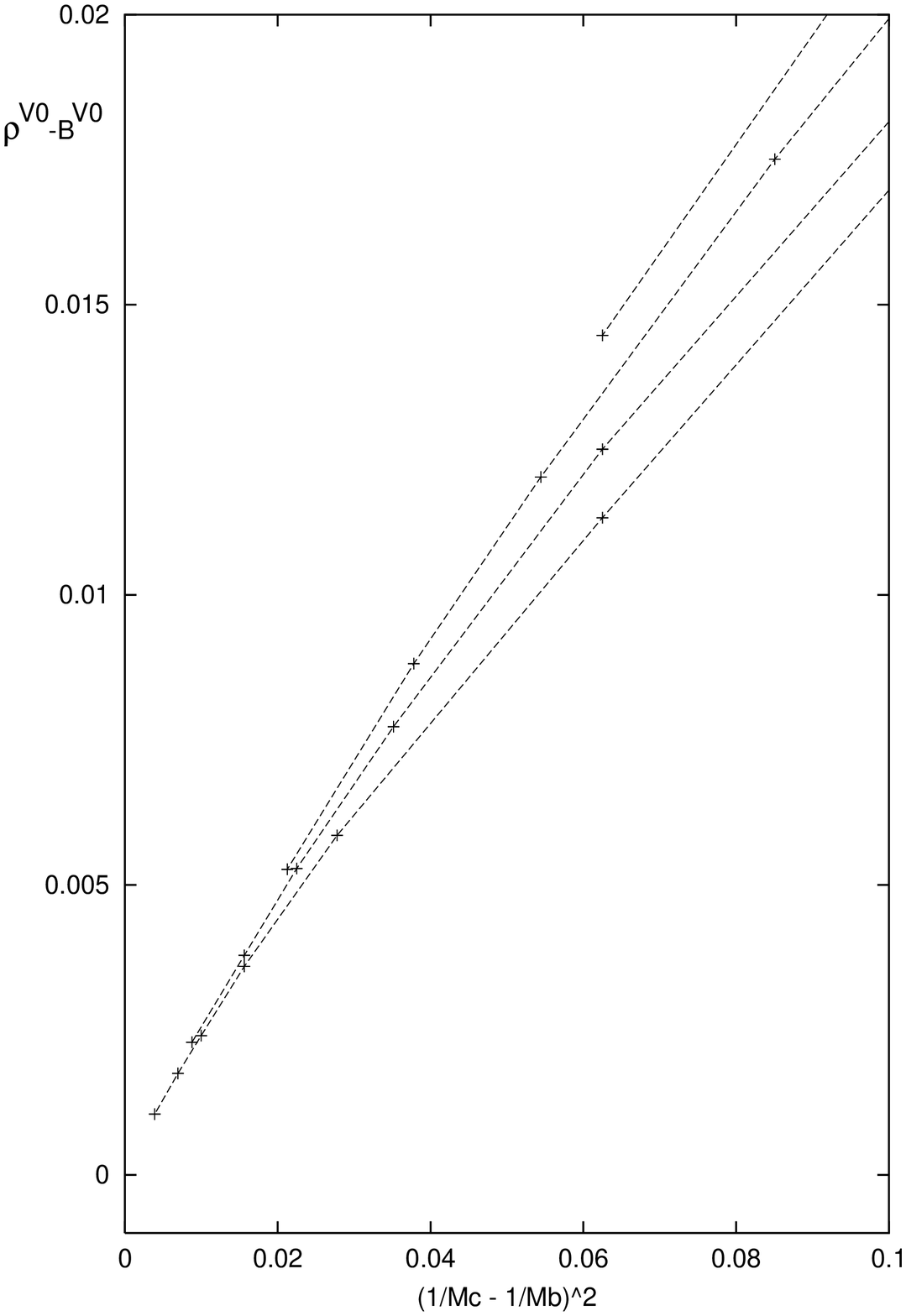}
\end{figure}

Luke's theorem \cite{Luke} shows that there is no $O(\frac{1}{M})$
correction to form factors involving the vector current in HQET at zero recoil. 
Luke's theorem is a non-perturbative
observation based upon the symmetries of HQET. The required
symmetries are satisfied by Lattice NRQCD, and the theorem should certainly be
manifested in the one-loop coefficients at zero-recoil. Demonstrating this
is less trivial than in HQET however, since $M$ dependence arises
out of terms in the action as well as through current insertions, and we
proceed by numerically calculating the $M$ dependence of the lattice NRQCD 
contribution to the renormalisation.
In figure~\ref{FIGluke} we display the lattice
one loop renormalisation  $\rho^{V_0} - B^{V_0}$ plotted versus $(\frac{1}{M_c} - \frac{1}{M_b})^2$,
for various values of $\xi = \frac{M_c}{M_b}$.
This plot shows that the leading correction is linear in $(\frac{1}{M_c} - \frac{1}{M_b})^2$ 
in the $\frac{1}{M}\rightarrow 0$ region, with a universal slope.
This is consistent with the absence of the $\frac{1}{M}$ term, 
as expected by Luke's theorem applied to lattice NRQCD.

Of course the analogue of Luke's theorem should be held non-perturbatively by lattice NRQCD,
and a fit of the dependence of the non-perturbative lattice form factor on $\frac{1}{M_b}$ near the
static limit should be made on simulation data to check the consistency with Luke's theorem.

\subsection{Spatial Axial Current}
\begin{figure}
\caption{$\rho^{A_k}$ as a function of $\xi$ and $\frac{1}{M_b}$.
\label{FIGrhoAk3d}}
\setlength\epsfxsize{250pt} \setlength\epsfysize{250pt}
\epsfbox{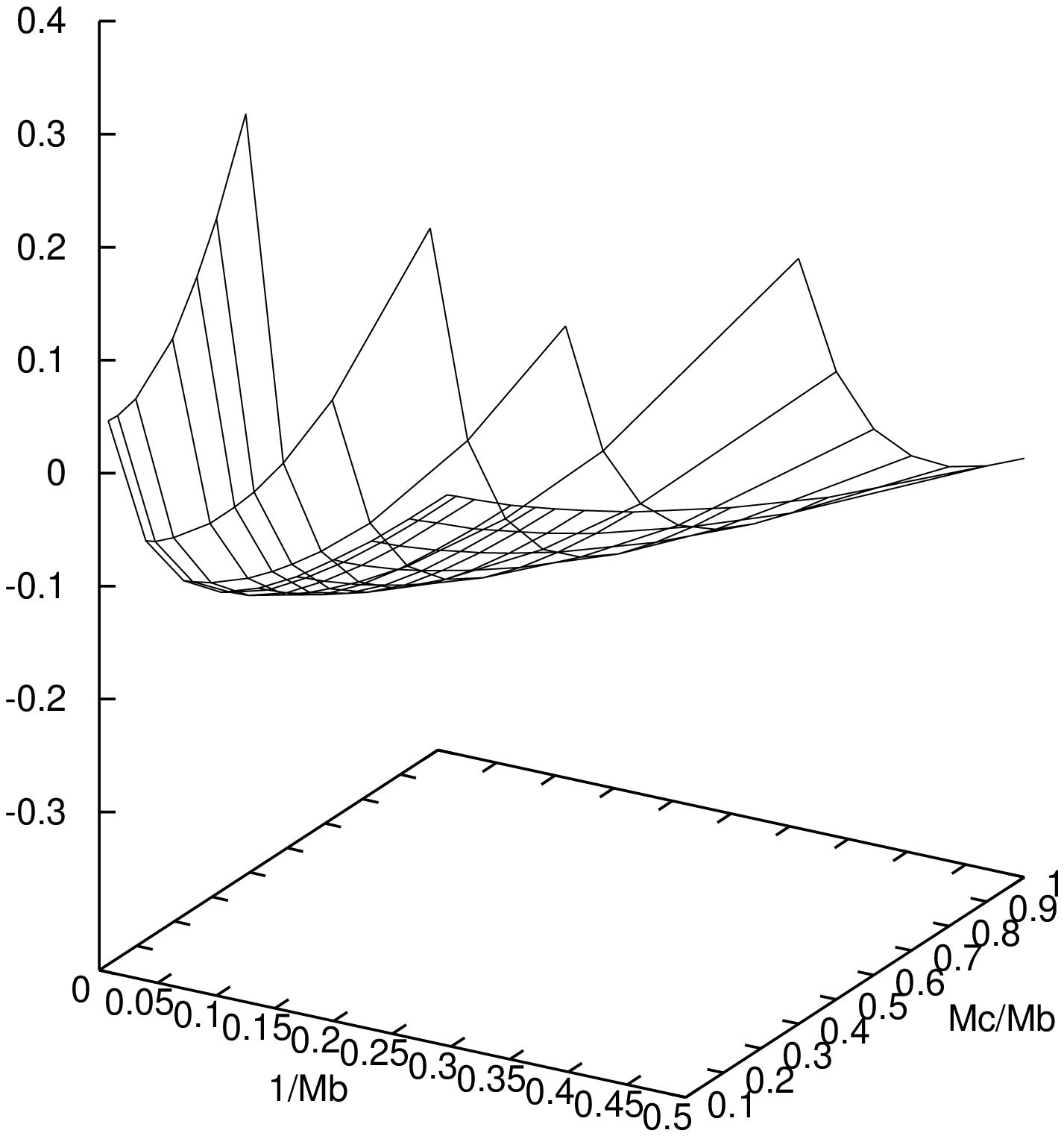}
\end{figure}

For the spatial axial current,
\begin{eqnarray}
A_k^{\overline{MS}} &=&
\left\{ 1 + \alpha_s \rho^{A_k}\right\} {\cal O}^{A_k}, \\
\rho^{A_k} &=& 
B^{A_k} - \half\{ {\bf Z}_{\psi_c}\oneloop + {\bf Z}_{\psi_b}\oneloop\}
- {\bf \Lambda}^{A_k \soneloop},\\
B^{A_k}&=& \frac{1}{3\pi}\left[ -3\frac{1+\xi}{1-\xi}\log\xi - 8\right].
\end{eqnarray}
We plot the renormalisation coefficient $\rho^{A_k}$ in figure~\ref{FIGrhoAk3d},
and give the correction as a function of $\frac{1}{M_b}$ for 
$\xi = 0.1,0.2,0.3,0.4,0.5$ and $ 1.0$ in table~\ref{TABrhoAk}.

\subsection{Temporal Axial Current}
\begin{figure}
\caption{$\rho_1^{A_0}$ as a function of $\xi$ and $\frac{1}{M_b}$.}
\label{FIGrhoA013d}
\setlength\epsfxsize{250pt} \setlength\epsfysize{250pt}
\epsfbox{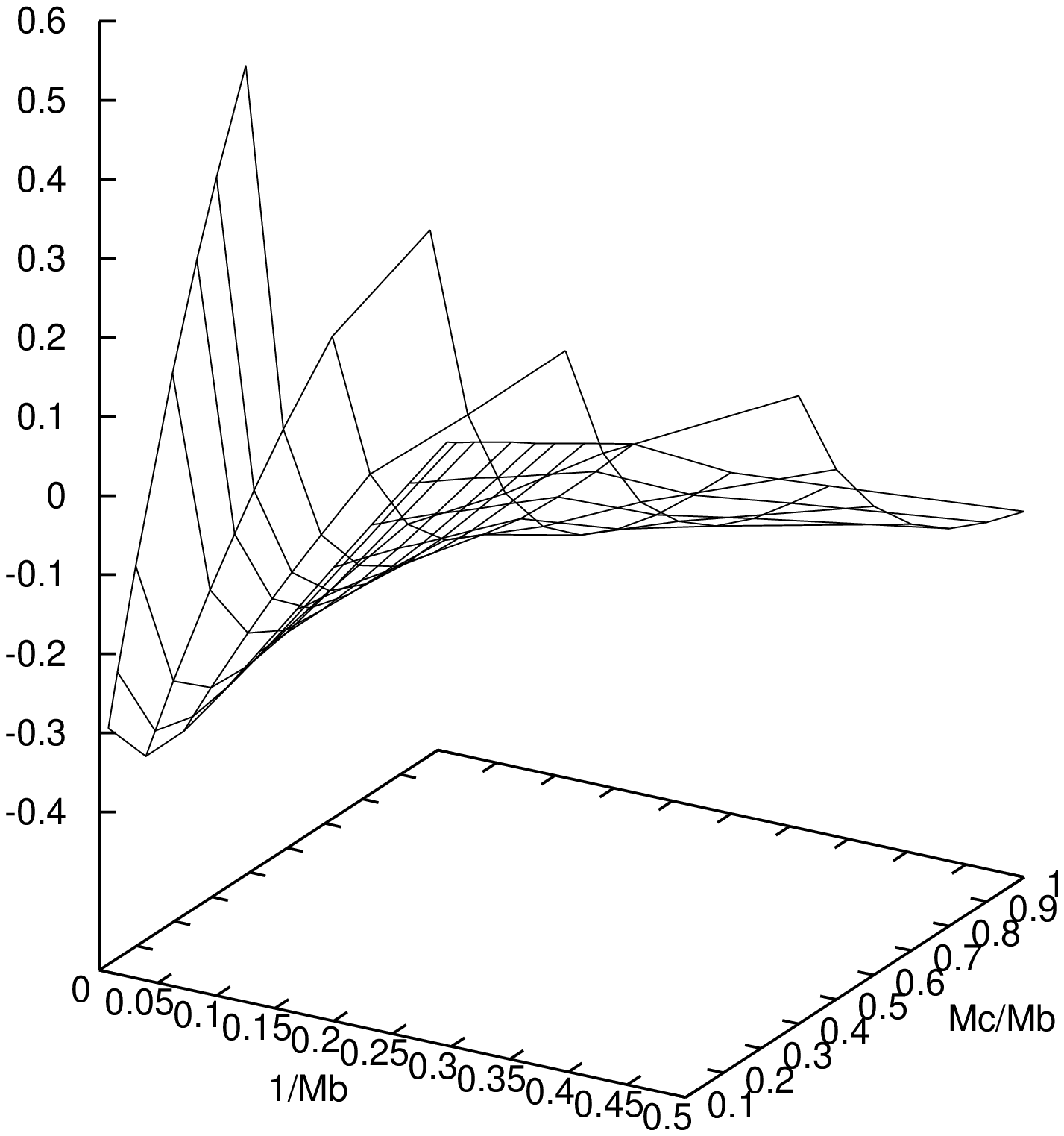}
\end{figure}

\begin{figure}
\caption{$\rho_2^{A_0}$ as a function of $\xi$ and $\frac{1}{m_b}$.}
\label{FIGrhoA023d}
\setlength\epsfxsize{250pt} \setlength\epsfysize{250pt}
\epsfbox{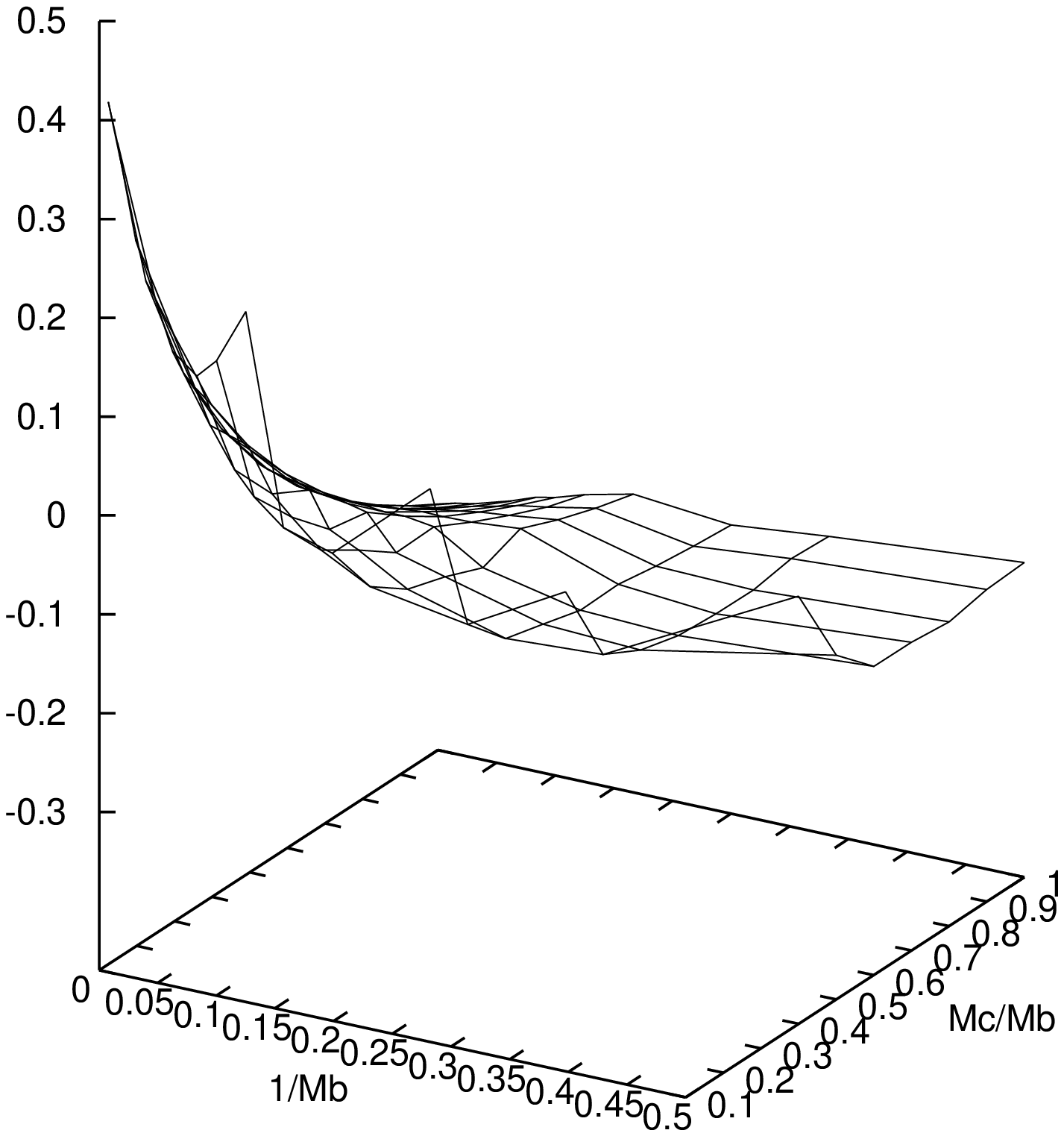}
\end{figure}

For the temporal axial current we have
\begin{equation}
A_0^{\overline{MS}} =
\left\{ 1 + \alpha_s \rho^{A_0}_1 \right\}{\cal O}_1^{A_0} +
\left\{ 1 + \alpha_s \rho^{A_0}_2 \right\}{\cal O}_2^{A_0} 
\end{equation}
where
\begin{eqnarray}
\rho^{A_0}_1 &=&
B_1^{A_0} - 
\half\{ {\bf Z}_{\psi_c}\oneloop + {\bf Z}_{\psi_b}\oneloop \}
-{\bf \Lambda}_{11}^{A_0 \soneloop}-{\bf \Lambda}_{21}^{A_0 \soneloop} -{\bf Z}_{m_b}\oneloop
-{\bf \Lambda}_{11}^{{\rm TI}^{A_0} \soneloop} -{\bf Z}_{m_b}^{{\rm TI} \soneloop} \\
\rho^{A_0}_2 &=&
B_2^{A_0} - 
\half\{ {\bf Z}\oneloop_{\psi_c} + {\bf Z}_{\psi_b}\oneloop \}
-{\bf \Lambda}_{12}^{A_0 \soneloop}-{\bf \Lambda}_{22}^{A_0 \soneloop}-{\bf Z}_{m_c}\oneloop
-{\bf \Lambda}_{22}^{{\rm TI}^{A_0} \soneloop}-{\bf Z}_{m_c}^{ {\rm TI} \soneloop}
\end{eqnarray}
and
\begin{eqnarray}
B^{A_0}_1 &=& \frac{1}{3\pi}
\left[-3\frac{1+\xi}{1-\xi}\log\xi-8 -6\frac{1+\xi}{1-\xi} - 12 \frac{\xi\log\xi}{(1-\xi)^2} \right] \\
B^{A_0}_2 &=&\frac{1}{3\pi}
\left[-3\frac{1+\xi}{1-\xi}\log\xi-8 +6\frac{1+\xi}{1-\xi} + 12 \frac{\xi\log\xi}{(1-\xi)^2} \right]
\end{eqnarray}
We plot the renormalisation coefficients $\rho^{A_0}_1$ and $\rho^{A_0}_2$ in 
figures~\ref{FIGrhoA013d} and \ref{FIGrhoA023d}.
The various contributions to $\rho_1^{A_0}$ for $\xi=0.3$ are given in table~\ref{TABrhoA0_1_contributions} 
allowing the reader to see relatively easily the change that must be applied
to obtain the coefficients with  a different tadpole improvement prescription, 
such as the mean link in Landau gauge.
The diagonal elements ${\bf \Lambda}_{\beta\beta}$ of the mixing matrix both contain infra-red divergences
which exactly cancel those of the wavefunction renormalisation, and have associated tadpole
improvement counter terms associated with the $u_0$ factor in the derivative operator.
It is, in fact, interesting to observe the tadpole improvement programme in action;
there are large cancellations between the 
${\bf Z}_m^{{\rm TI} \soneloop}$ and ${\bf Z}_m\oneloop$ terms and between
the $\Lambda_{11}^{{\rm TI}^{A_0} \soneloop}$ and the 
$\half\{{\bf Z}_{\psi_c}\oneloop + 
{\bf Z}_{\psi_b}\oneloop\} + {\bf \Lambda}^{A_0 \soneloop}_{11}$ terms.
This cancellation is even more manifest when we isolate 
the tadpole graph contributions within ${\bf \Lambda}^{A_0 \soneloop}_{11}$
and ${\bf Z}_m\oneloop$.
The two overall correction coefficients are tabulated as a function of $\frac{1}{M_b}$ for 
$\xi = 0.1,0.2,0.3,0.4,0.5$ and $1.0$ in tables~\ref{TABA0rho1} and \ref{TABA0rho2}.

\subsection{Spatial Vector Current}
\begin{figure}
\caption{$\rho_1^{V_k}$ as a function of $\xi$ and $\frac{1}{m_b}$ .}
\label{FIGrhoVk13d}
\setlength\epsfxsize{250pt} \setlength\epsfysize{250pt}
\epsfbox{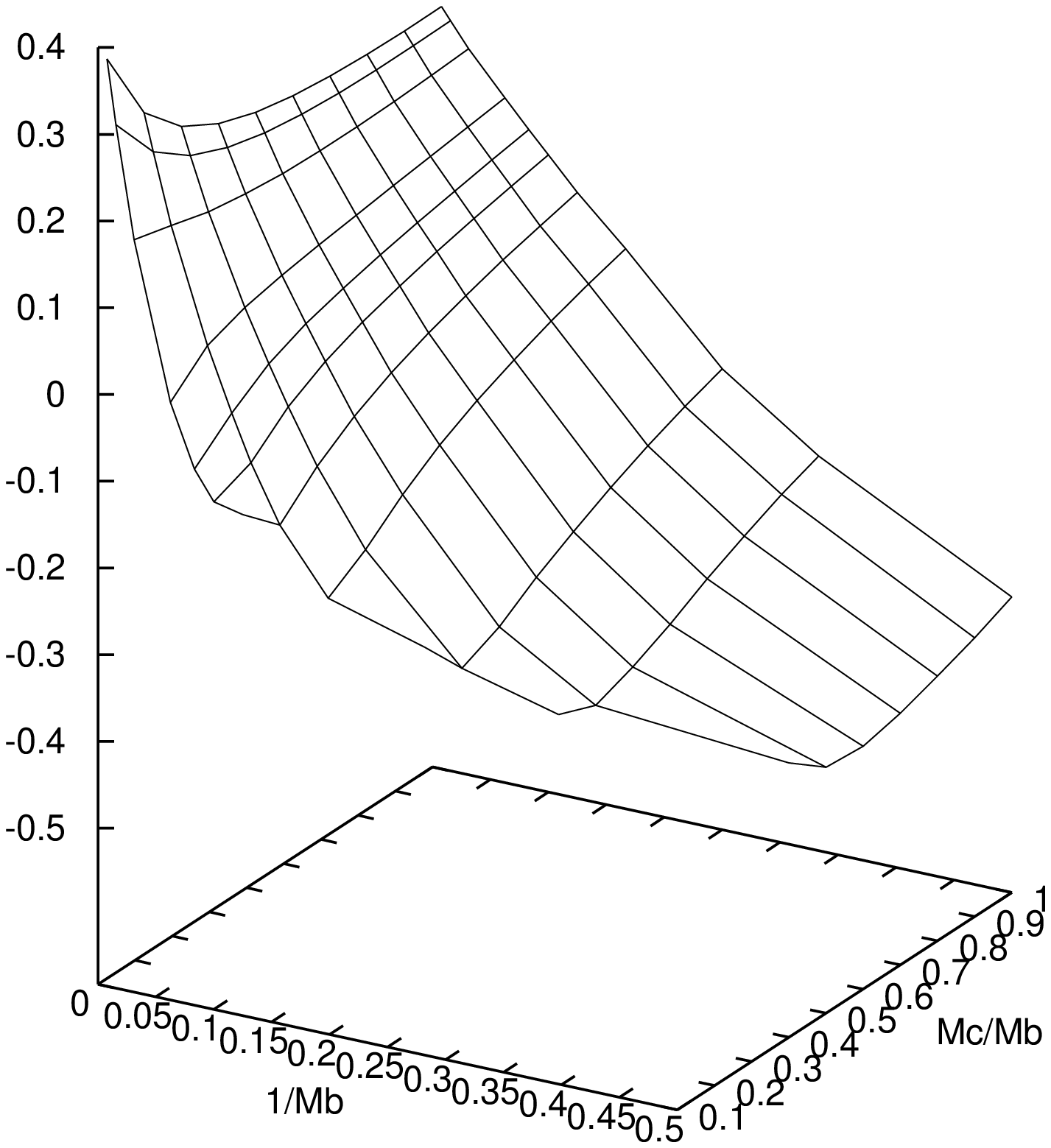}
\end{figure}

\begin{figure}
\caption{$\rho_2^{V_k}$ as a function of $\xi$ and $\frac{1}{m_b}$.}
\label{FIGrhoVk23d}
\setlength\epsfxsize{250pt} \setlength\epsfysize{250pt}
\epsfbox{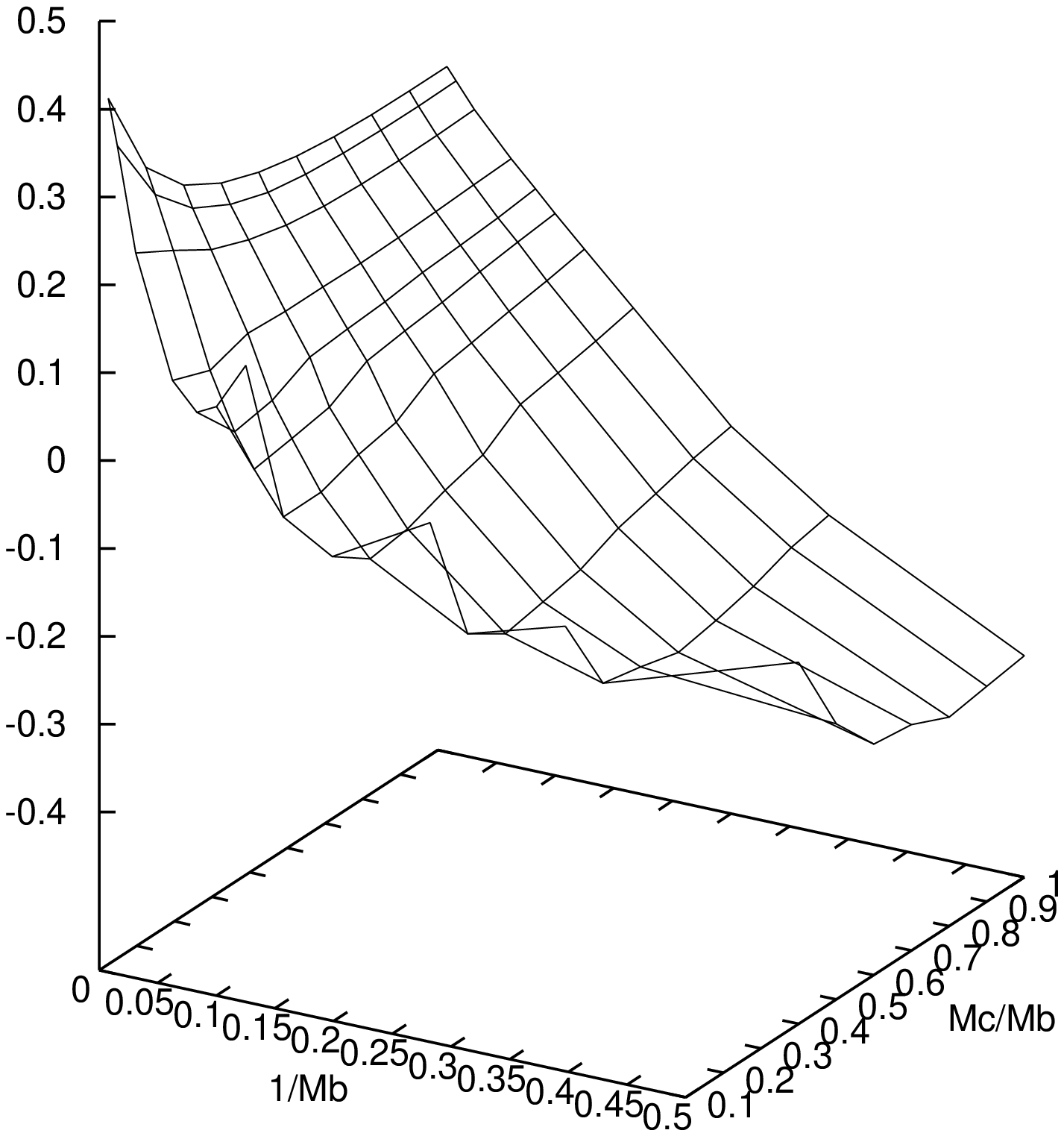}
\end{figure}

\begin{figure}
\caption{$\rho_3^{V_k}$ as a function of $\xi$ and $\frac{1}{m_b}$.}
\label{FIGrhoVk33d}
\setlength\epsfxsize{250pt} \setlength\epsfysize{250pt}
\epsfbox{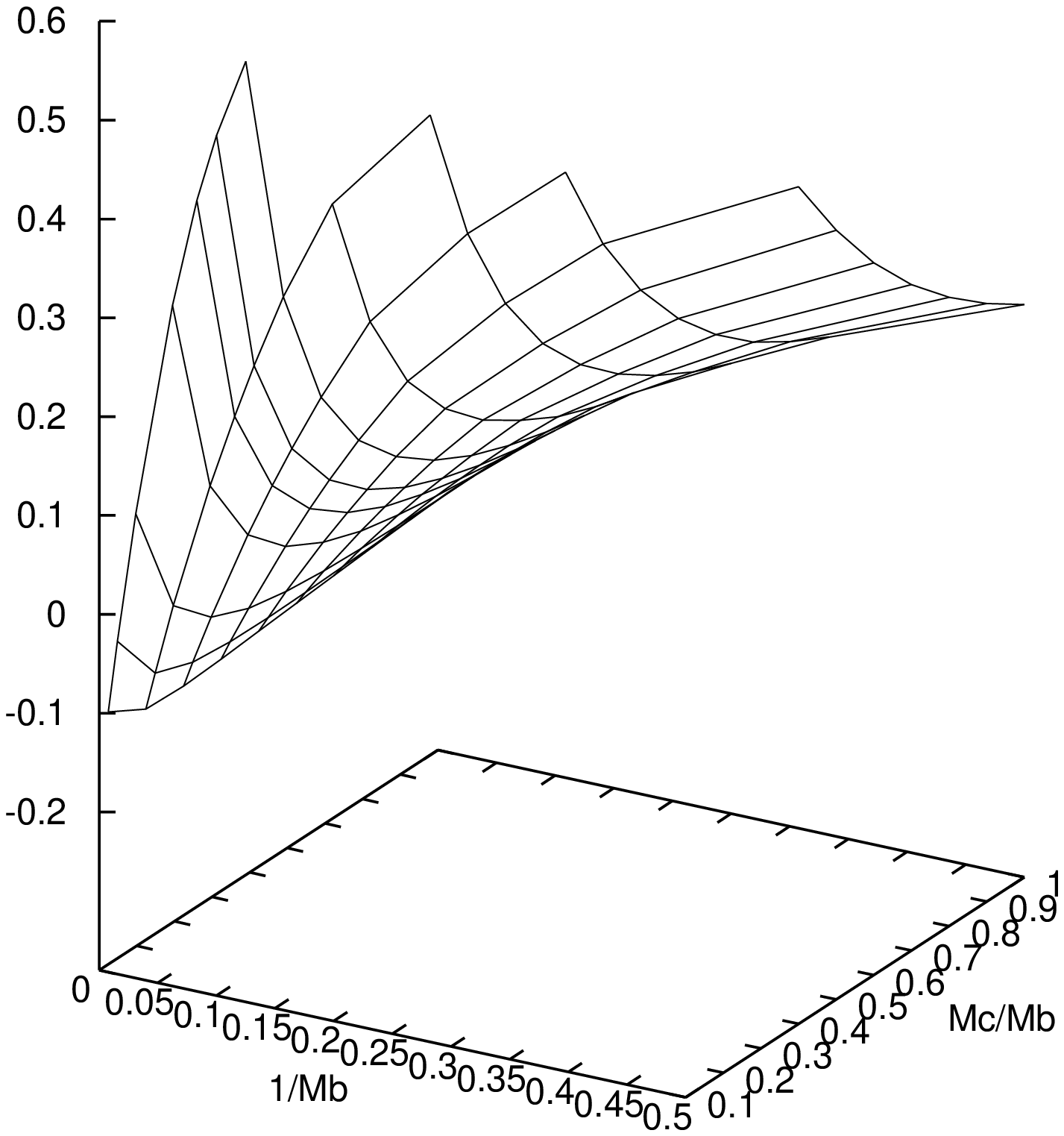}
\end{figure}

\begin{figure}
\caption{$\rho_4^{V_k}$ as a function of $\xi$ and $\frac{1}{m_b}$.}
\label{FIGrhoVk43d}
\setlength\epsfxsize{250pt} \setlength\epsfysize{250pt}
\epsfbox{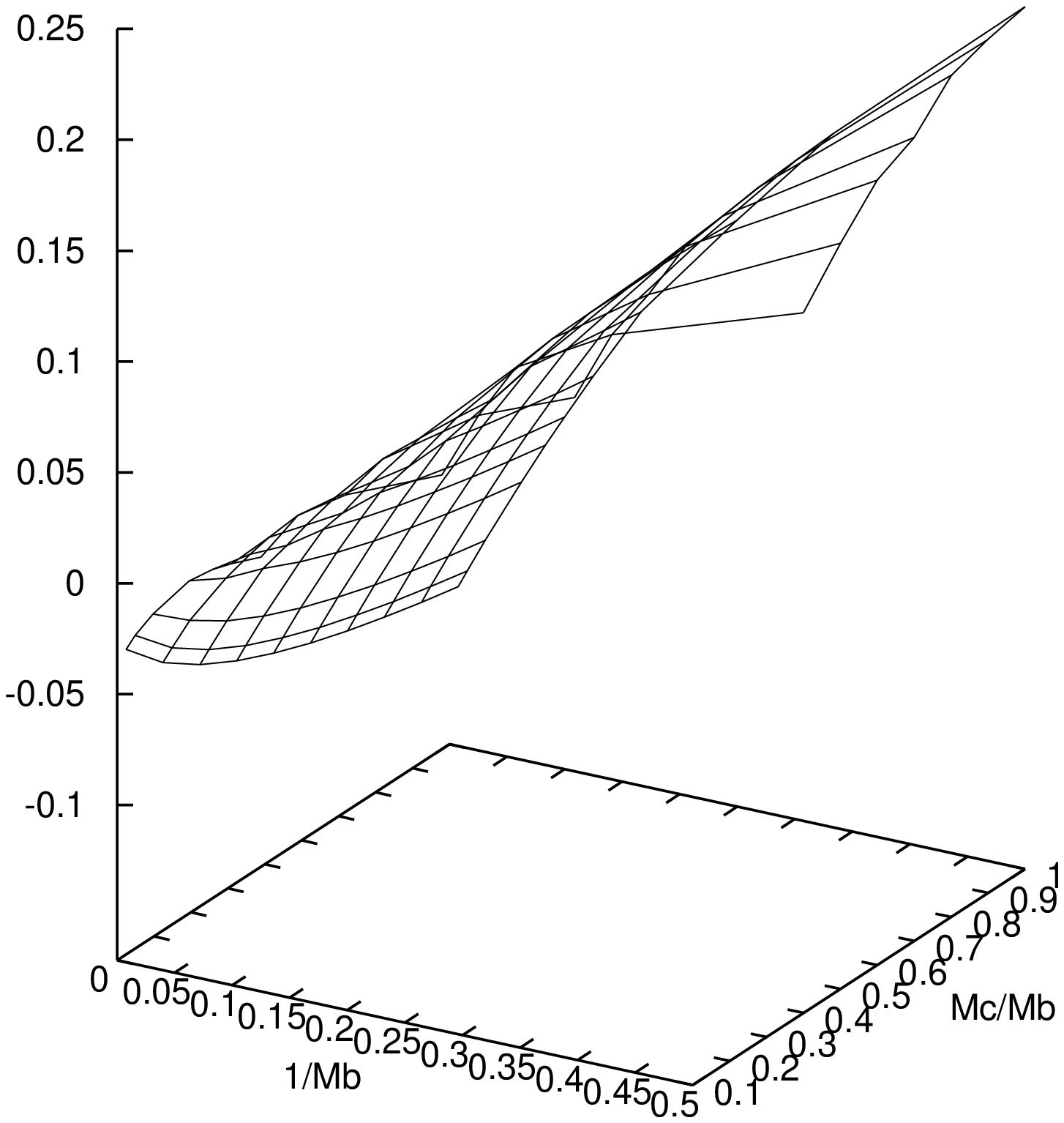}
\end{figure}

For the spatial vector current we obtain
\begin{equation}
\begin{array}{ccc}
V_k^{\overline{MS}}  &=&
\left\{ 1 + \alpha_s \rho^{V_k}_1\right\} {\cal O}_1^{V_k} 
+\left\{ 1 + \alpha_s \rho^{V_k}_2\right\} {\cal O}_2^{V_k} 
+\alpha_s \rho^{V_k}_3 {\cal O}_3^{V_k} 
+\alpha_s \rho^{V_k}_4 {\cal O}_4^{V_k} 
\end{array}
\end{equation}
where
\begin{eqnarray}
\rho^{V_k}_1 &=&  B_1^{V_k} - 
\half\{ {\bf Z}_{\psi_c}\oneloop + {\bf Z}_{\psi_b}\oneloop\}
-{\bf \Lambda}_{11}^{V_k \soneloop}-{\bf \Lambda}_{21}^{V_k \soneloop} -{\bf Z}_{m_b}\oneloop  
-{\bf \Lambda}_{11}^{{\rm TI}^{V_k} \soneloop} -{\bf Z}_{m_b}^{{\rm TI} \soneloop} \\
\rho^{V_k}_2&=& B_2^{V_k} - 
\half\{ {\bf Z}_{\psi_c}\oneloop + {\bf Z}_{\psi_b}\oneloop\}
-{\bf \Lambda}_{12}^{V_k \soneloop}-{\bf \Lambda}_{22}^{V_k \soneloop}- {\bf Z}_{m_c}\oneloop
-{\bf \Lambda}^{{\rm TI}^{V_k} \soneloop}_{22} -{\bf Z}_{m_c}^{{\rm TI} \soneloop} \\
\rho^{V_k}_3 &=& B_3^{V_k} 
-{\bf \Lambda}_{13}^{V_k \soneloop}-{\bf \Lambda}_{23}^{V_k \soneloop}\\
\rho^{V_k}_4 &=&  B_4^{V_k} 
-{\bf \Lambda}_{14}^{V_k \soneloop}-{\bf \Lambda}_{24}^{V_k \soneloop}\\
\end{eqnarray}
and
\begin{eqnarray}
B_1 ^{V_k}&=& B_2^{V_k} = \frac{1}{3\pi}\left[-3\frac{1+\xi}{1-\xi}\log\xi - 4 \right]\\
B_3^{V_k} &=&\frac{1}{3\pi}\left[ -\frac{2}{1-\xi} - \frac{2\xi\log\xi}{(1-\xi)^2}\right]\\
B_4^{V_k} &=&\frac{1}{3\pi}\left[ \frac{2\xi}{1-\xi}+\frac{2\xi\log\xi}{(1-\xi)^2}\right].
\end{eqnarray}
Since the ${\cal O}_3$ and ${\cal O}_4$ do not appear at tree level, we can see that it is
in fact unnecessary to calculate ${\bf \Lambda}_{3\beta}^{V_k}$ and ${\bf \Lambda}_{4\beta}^{V_k}$.
We plot the renormalisation coefficients $\rho^{V_k}_1\ldots\rho^{V_k}_4$ in figures~\ref{FIGrhoVk13d}
to \ref{FIGrhoVk43d}.
The various contributions to $\rho_1^{V_k}$ and $\rho_3^{V_k}$ are given for $\xi=0.3$ in table~\ref{TABrho1Vk_contributions}
and table~\ref{TABrho3Vk_contributions}.
The four overall correction coefficients are given as a function of $\frac{1}{M_b}$ for 
$\xi = 0.1,0.2,0.3,0.4,0.5$ an $1.0$ in tables~\ref{TABVkrho1} to \ref{TABVkrho4}.

\section{Relevant Momentum Scale for $\alpha_s$}
\label{SecScale}
In section~\ref{SecIntro} we argued that the regularisation scheme correction
is dominated by ultra-violet physics and is therefore calculable within perturbation
theory. This is addressed in this section by attempting to calculate the appropriate scale
for $\alpha_s$ in the correction to $V_0$ and $A_k$. 

The Lepage-Mackenzie $q^\ast$ scale \cite{LepageMackenzie} 
for some one-loop process whose Feynman amplitude $A = \intk I(k)$ is defined by
\begin{equation}
\label{LMscale}
\log (a q^{\ast})^2 = \frac{\intk I(k) \log (a k)^2}{\intk I(k)}.
\end{equation}
When a one-loop expression is made up of two sub-processes, $A = \intk I_A(k)$ and $B = \intk I_B(k)$ 
we may combine the $q^\ast$'s as follows,
\begin{equation}
\label{CombineQs}
\log (aq^{\ast})^2_{A+B} = \frac{\intk I_A(k) \log(ak)^2 + \intk I_B(k) \log(ak)^2}{\intk I_A(k)  + \intk(I_B(k)}
\end{equation}
or more conveniently,
\begin{equation}
q^\ast_{A+B} = \left[(q^\ast_A)^A \times (q^\ast_B)^B \right]^{\frac{1}{A+B}}.
\end{equation}
Here we shall calculate the Lepage-Mackenzie $q^\ast$ from lattice perturbation theory, and then
use this rule to combine it with the continuum BLM scales from section \ref{ContBLM}
to obtain an overall scale. 

The combination of lattice graphs containing the cancelling IR divergences proved troublesome using 
4 dimensional numerical integration, so we used the (non-rigorous) 
Hernandez and Hill prescription \cite{HernandezHill}
to deal with the $k_0$ integral, and integrated the remaining spatial momenta numerically.

We present the results for $q^\ast_{V_0}$ in table~\ref{TABqastV0}.
The overall scale is reassuringly large, and is dominated by the 
continuum scale due to the typical relative size of the corrections.
We therefore believe that the use of Hernandez and Hill prescription to 
obtain numerical stability constitutes an irrelevant error in the overall scale. 
We note that $\alpha_V(q^\ast) \le 0.22$ for $q^\ast \ge 3 GeV$, which is 
the case in a typical simulation regimes for $V_0$ \cite{Joachim}.

We present the results  for $q^\ast_{A_k}$ in table~\ref{TABqastAk}.
There is a line (straddled by the bold highlighted data in table~\ref{TABqastAk} )
in the $\xi - M_b$ plane where $q^\ast$ 
becomes poorly behaved. 
It can be seen in table~\ref{TABqastAkcontrib} that the individual scales 
$q_{A_k}^{\ast {\rm lat}} $ and $q^{\ast {\rm cont}}_{A_k}$ are large at this point, 
but that the overall one-loop correction is near vanishing. The location
of this uncontrolled behaviour is not dependent on our use of the Hernandez and Hill
prescription, however the details of the  behaviour will, of course, be somewhat
dependent. The zero in the overall correction introduces a 
vertical asymptote for $\log (a q^*)^2$, due to the vanishing denominator in (\ref{CombineQs})
and consequently a limit of infinity or zero for $q^\ast$ depending on the direction
of approach. 
The resultant low scale on one side of the zero in the correction is not considered problematic
by the authors, since it is a breakdown of the mean scale interpretation of (\ref{LMscale}), that
occurs whenever the correction is small.\footnote{As noted by Morningstar \cite{ColinMassQstar}, the 
integrand $I_A(k) + I_B(k)$ must be either strictly positive or strictly negative for the mean
value theorem to guarantee $0\le a q^\ast \le 2\pi$. Exact cancellation of the
integrals is likely to exacerbate the problem.
}

The prescription used to 
set the scale at which to evaluate $\as$  in a mixing calculation
is considered an open question by the authors, however we remain 
encouraged by the reliability of the perturbation theory suggested by
the scales determined thus far. 

\section{Summary}
We have shown how to construct the $\overline{MS}$ weak decay currents for 
$B\rightarrow D^{(\ast)}$ transitions
from operators in the lattice NRQCD effective theory at and around the zero recoil point for a wide range
of phenomenologically relevant lattice mass parameters.
These constants are of use to both recent \cite{Joachim} and future \cite{FutureBtoD} calculations of the
semileptonic decay form factors. The one-loop coefficients typically have modulus 
$\le 0.3$ in the usual simulation regime.
The lattice temporal vector current is not renormalised in the 
degenerate quark case
(this is not the case for Wilson fermions),
with the result that
the current correction for physical $\frac{m_c}{m_b}$ ratios is particularly small.
For two of the currents we have estimated the Lepage-Mackenzie $q^\ast$ scales, 
and find that it is large, giving
overall one-loop corrections of order $\le 5$\%, and suggesting perturbation
theory is converging adequately. With further suppression, the
next (and higher) order corrections in the operator
renormalisation could well be below the expected statistical error of the BaBar data.
\footnote{
Certainly, the two loop renormalisation correction to HQET operators is known to be $O(1\%)$ 
\cite{TwoLoopHQET}.
}

\section{Acknowledgments}
    We wish to thank Junko Shigemitsu for many useful conversations, and 
    cross checking the mass renormalisation. 
    We made use of unpublished notes by Colin Morningstar for checking our continuum 
    calculation.
    Peter Boyle is supported by PPARC grant PP/CBA/62 . 
    Both Peter Boyle and Christine Davies fondly acknowledge the hospitality of the 
    UCSB, Santa Barbara where some of this work was carried out.

\begin{appendix}
\section{Self Energy Calculation}
\label{AppA}
\begin{figure}[htb]
\caption{Regular topology self energy contribution. \\
Contributions arise from a temporally polarised gluon coupling to 
vertices arising from the $\partial_t$
term in the action, and spatially polarised
gluons coupling to vertices arising from the $\frac{p^2}{2m}$ term 
and the $\frac{\sigma\cdot B}{2m}$ term.
For the wavefunction renormalisation the mixed $\frac{\sigma\cdot B}{2m}$ and $\frac{p^2}{2m}$
coupling diagram vanishes at $\vec{p}=0$, but, since the derivatives do not, the
mixed diagrams contribute to the kinetic mass renormalisation.}
\label{FIGself_reg}
\setlength\epsfxsize{150pt}
\epsfbox{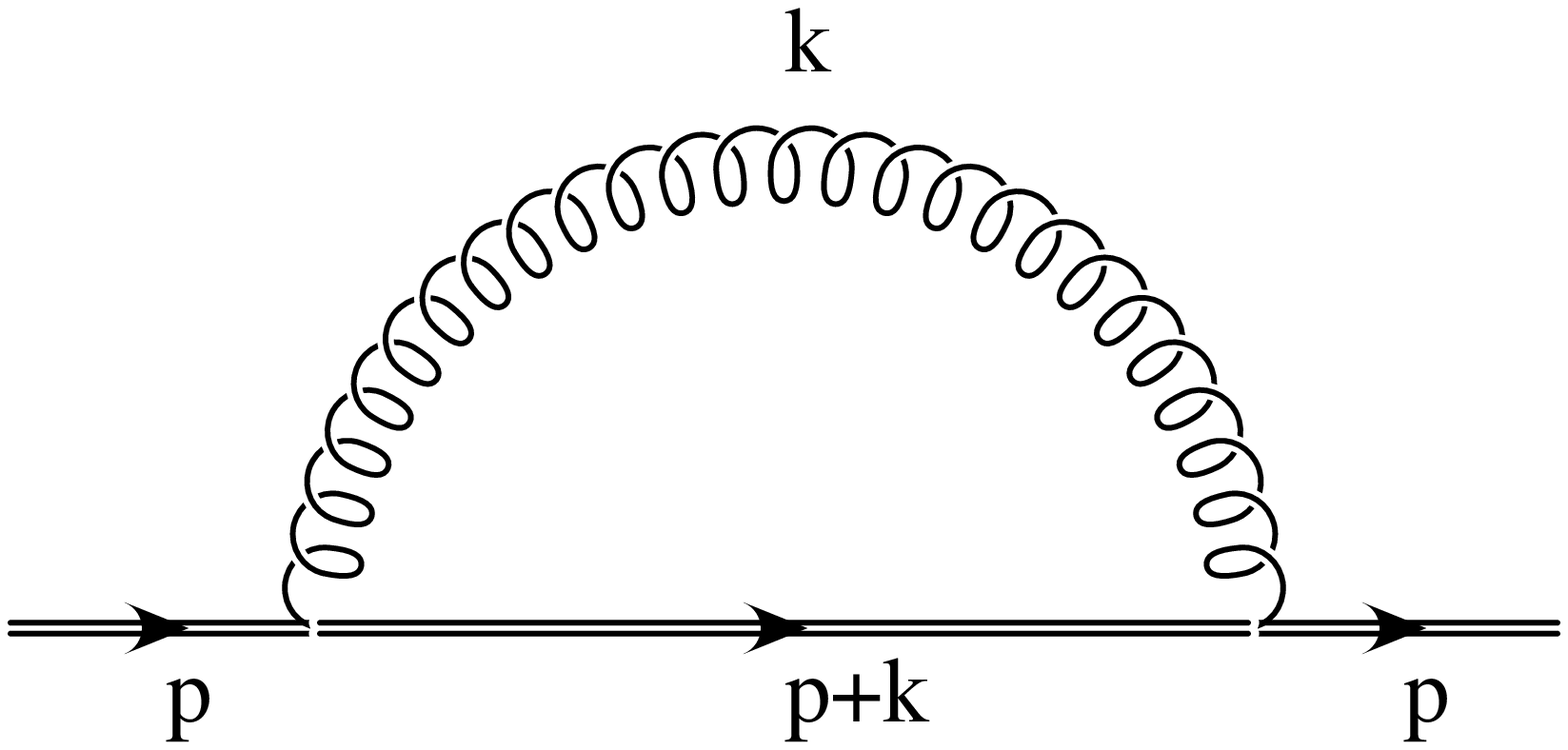}
\end{figure}

\begin{figure}[htb]
\caption{Tadpole contribution to the self energy.\\
Due to the external equation of motion, this diagram does not in fact contribute
to the wavefunction renormalisation, however it does contribute to the
kinetic mass renormalisation, through both $\frac{p^2}{2m}$ and $\frac{\sigma\cdot B}{2m}$
two gluon couplings.}
\label{FIGself_tad}
\setlength\epsfxsize{150pt}
\epsfbox{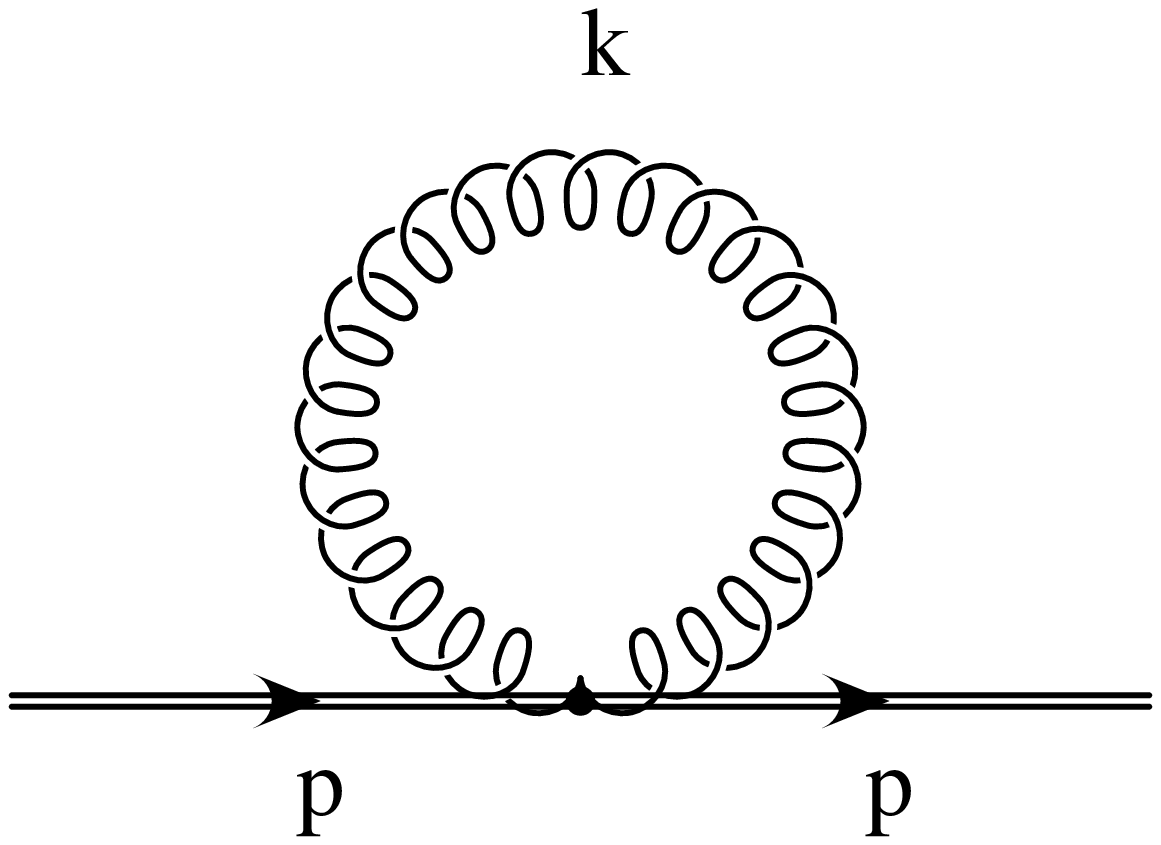}
\end{figure}

To one loop the  graph topologies  shown in figures~\ref{FIGself_reg} and \ref{FIGself_tad} 
contribute to the lattice self energy. The diagrams contain contributions
from both the temporally polarised gluon vertex, and from spatially polarised gluons
via the $\frac{p^2}{2m}$ term and via the $\sigma \cdot B$ term.
The lattice self energy must satisfy cubic invariance in the
spatial momenta, and we write the first terms of a double taylor series for the
self-energy as 
\begin{equation}
\Sigma^{\rm lat}(a p_0,a \vec{p}) = g^2 \left\{ \Omega_0 + \Omega_1 i a p_0 + \Omega_2 \frac{p^2 a}{2M_0 a} + \ldots\right\}
\end{equation}
where clearly,
\begin{eqnarray}
g^2 \Omega_0 &=& \Sigma(0,0)\\
g^2 \Omega_1 &=& -i\frac{\partial}{\partial a p_0}\Sigma(0,0)\\ 
g^2 \Omega_2 &=& a M_0 \frac{1}{a^2} \frac{\partial^2}{\partial p_j^2} \Sigma(0,0)\\ 
\end{eqnarray}
to order $\alpha$ then the small $\frac{v^2}{c^2}$ expansion of the heavy quark propagator is 
\begin{eqnarray} 
G^{-1}_r(p) &=& G^{-1}(p) - \Sigma(a p_0,a \vec{p})\\
            &=& ip_0 + \frac{p_0^2}{2} + \frac{p^2}{2M_0} -ip_0 \frac{p^2}{2M_0} 
                - g^2\Omega_0 - i a p_0 g^2\Omega_1 - \frac{p^2}{2M_0} g^2 \Omega_2\\
            &=& Z_\psi ( -i a\bar{p}_0 + \frac{p^2a^2}{2Z_m^{\rm kin}Ma}  + \frac{\bar{p}_0^2 a^2}{2} )
\end{eqnarray}
where to order $g^2$ and $\frac{v^2}{c^2}$ we have,
\begin{eqnarray}
i \bar{p}_0 &=& i p_0 - g^2 \Omega_0 \\
Z_\psi    &=& 1 - g^2(\Omega_0+\Omega_1) \\
Z_m^{\rm kin} &=& 1 + g^2 (\Omega_2 - \Omega_1)
\end{eqnarray}
The contributions from the tadpole diagram figure~\ref{FIGself_tad}
to $\Omega_0$ and $\Omega_1$ explicitly cancel in $Z_\psi$ due to the equation
of motion of the external spinor. 
In the $Z_m$ calculation, both the tadpole and normal topologies contribute.
In order to compute $Z_m$ it was necessary to
code a routine to compute the second derivative of the self energy with respect to one of the external
spatial momentum components. This was fairly involved, due to the nature of the
lattice Feynman rules. There are cancelling infra-red divergences in both $\Omega_0$ and
in $\Omega_1$, so that the integral over the $k_0$ was performed analytically and the 
remaining three dimensions integrated numerically.
Our numbers for both the $Z_\psi$, $\Delta_E$, and $Z_m$ agree within statistical
error with those of Morningstar and Shigemitsu \cite{JunkoColin}.

\end{appendix}
\pagebreak
 

\begin{figure}[htb]
\caption{Diagrams of this topology contribute to both the temporal axial and
spatial vector currents. The ''tadpole'' arises from gauge link
in the derivative in the current operators.
}
\label{FIGvertex_tad}
\setlength\epsfxsize{150pt}
\epsfbox{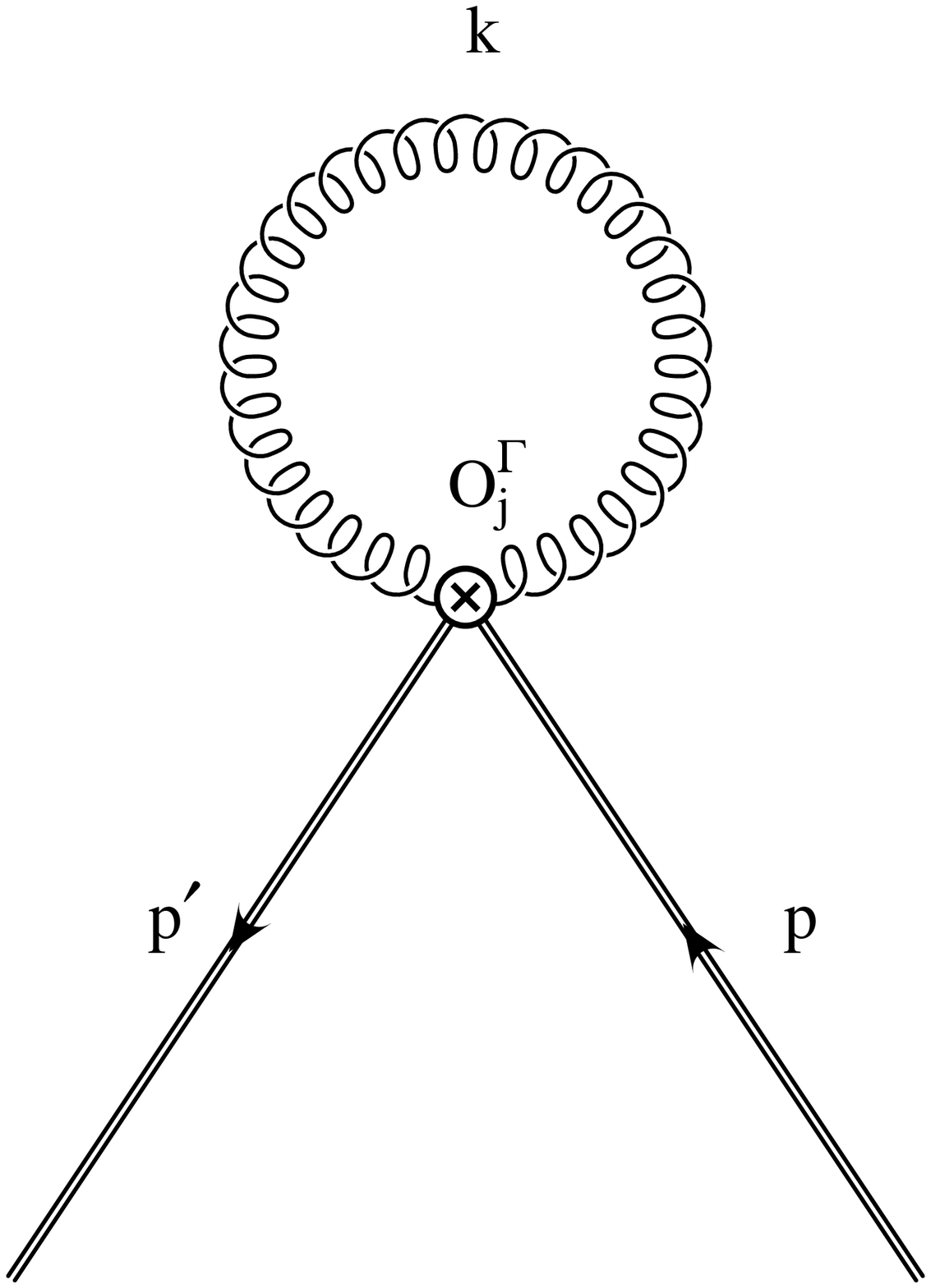}
\end{figure}

\begin{table}[hbt]
\caption{
\label{TabContOps}
Continuum operators for which we will select corresponding Lattice operators.
}
\begin{tabular}{c|c|c|c}
$V_0$ & $A_k$ & $V_k$ & $A_0$ \\
\hline
$\Omega_1^{V_0} = 1_{\rm pauli}$ & $\Omega_1^{A_k} = -i \sigma_k $ 
&
$\begin{array}{ccc}
\Omega_1^{V_k} &=& -i \sigma_k \frac{\sigma \cdot p}{2m_b}\\
\Omega_2^{V_k} &=& -i \frac{\sigma \cdot p^\prime}{2m_c} \sigma_k\\
\Omega_3^{V_k} &=& -i \frac{p_k}{m_b} \\
\Omega_4^{V_k} &=& -i \frac{p^\prime_k}{m_c} 
\end{array}
$ 
&
$
\begin{array}{ccc}
\Omega_1^{A_0} &=&  \frac{\sigma \cdot p}{2m_b}\\
\Omega_2^{A_0} &=&  \frac{\sigma \cdot p^\prime}{2m_c} 
\end{array}
$
\end{tabular}
\end{table}

\begin{table}[hbt]
\caption{
\label{TabLatOps}
Bases of Lattice Currents to be matched to each of the weak currents.
}
\begin{tabular}{c|c|c|c}
$V_0$ & $A_k$ & $V_k$ & $A_0$ \\
\hline
${\cal O}_1^{V_0} = 1_{\rm pauli}$ & ${\cal O}_1^{A_k} = -i \sigma_k $ 
&
$\begin{array}{ccc}
{\cal O}_1^{V_k} &=& - \sigma_k \frac{\sigma \cdot \dr}{2M_b}\\
{\cal O}_2^{V_k} &=&  \frac{\sigma \cdot \dl}{2M_c} \sigma_k\\
{\cal O}_3^{V_k} &=& - \frac{\dr_k}{M_b} \\
{\cal O}_4^{V_k} &=& \frac{\dl_k}{M_c} 
\end{array}
$ 
&
$
\begin{array}{ccc}
{\cal O}_1^{A_0} &=& -i \frac{\sigma \cdot \dr}{2M_b}\\
{\cal O}_2^{A_0} &=&  i \frac{\sigma \cdot \dl}{2M_c} 
\end{array}
$
\end{tabular}
\end{table}

\begin{table}[hbt]
\caption{
\label{TABnhams}
Selected stabilisation parameter in different mass regions.
}
\begin{tabular}{c|c|c|c|c}
$M \ge 0.8 $ &$ M\ge 1.0$ &$M \ge 1.2$ &$M\ge 1.6$ & $M > 4.0$\\
\hline
 n = 5 & n=4 & n=3 & n=2 & n=1
\end{tabular}
\end{table}

%
%
\begin{table}[hbt]
\caption{
\label{TABrhoV0}
The one loop piece of the temporal vector current renormalisation for $\xi = 0.1,0.2,0.3,0.4,0.5,$ and $1.0$ .}
\begin{tabular}{c|c|c|c|c|c|c}
$m_b$ &$ \rho^{A_k}(\xi = 0.1) $ &$ \rho^{A_k}(\xi = 0.2) $ &$ \rho^{A_k}(\xi = 0.3) $ &$ \rho^{A_k}(\xi = 0.4) $ &$ \rho^{A_k}(\xi = 0.5) $ &$ \rho^{A_k}(\xi = 1.0) $ \\
\hline
2.000  & -& -& -& 0.127146(7)& 0.063266(3)& 0\\
3.000  & -& -& 0.168931(8)& 0.084085(3)& 0.043936(2)& 0\\
4.000  & -& 0.28746(1)& 0.131438(5)& 0.068204(2)& 0.0366150(9)& 0\\
6.000  & -& 0.208433(6)& 0.103350(2)& 0.0564405(10)& 0.0311405(5)& 0\\
8.000  & 0.46550(2)& 0.178367(4)& 0.092615(1)& 0.0516583(6)& 0.0288833(3)& 0\\
10.000  & 0.39874(1)& 0.163900(3)& 0.0871368(10)& 0.0492116(4)& 0.0276885(2)& 0\\
12.000  & 0.360656(8)& 0.155447(2)& 0.0839188(7)& 0.0477178(3)& 0.0270377(1)& 0\\
16.000  & 0.320742(5)& 0.146303(1)& 0.0803731(4)& 0.0462188(2)& 0.02633675(8)&0 
\end{tabular}

\end{table}

%
%
\begin{table}[hbt]
\caption{
\label{TABrhoAk}
The one loop contribution to the spatial axial current renormalisation for $\xi = 0.1,0.2,0.3,0.4,0.5$ and $1.0$ .}
\begin{tabular}{c|c|c|c|c|c|c}
$m_b$ &$ \rho^{A_k}(\xi = 0.1) $ &$ \rho^{A_k}(\xi = 0.2) $ &$ \rho^{A_k}(\xi = 0.3) $ &$ \rho^{A_k}(\xi = 0.4) $ &$ \rho^{A_k}(\xi = 0.5) $ &$ \rho^{A_k}(\xi = 1.0) $ \\
\hline
2.000  & -& -& -& 0.23875(1)& 0.117205(8)& -0.066846(4)\\
3.000  & -& -& 0.161977(8)& 0.031141(5)& -0.036802(4)& -0.140394(2)\\
4.000  & -& 0.251373(9)& 0.042554(5)& -0.048200(3)& -0.096478(2)& -0.169133(1)\\
6.000  & -& 0.081220(4)& -0.049163(2)& -0.109181(2)& -0.142730(1)& -0.1917372(5)\\
8.000  & 0.346311(8)& 0.016752(3)& -0.083899(1)& -0.1327985(9)& -0.1605718(7)& -0.2001877(3)\\
10.000  & 0.248523(5)& -0.014434(2)& -0.1012549(9)& -0.1445488(6)& -0.1695061(4)& -0.2043054(2)\\
12.000  & 0.192853(4)& -0.032424(1)& -0.1112436(7)& -0.1513905(4)& -0.1744741(3)& -0.2066181(1)\\
16.000  & 0.134707(2)& -0.0515471(7)& -0.1218945(4)& -0.1583394(3)& -0.1796509(2)& -0.20898845(8)
\end{tabular}

\end{table}

%
%
\begin{table}[hbt]
\caption{Contributions to the temporal axial current renormalisation for $\xi = 0.3$ .
\label{TABrhoA0_1_contributions}}
\begin{tabular}{c|c|c|c|c|c|c|c|c|c|c}
$m_b$ & $n_b$& $m_c$ & $n_c$ & $\half\{Z_{\psi_c} + Z_{\psi_b}\} + \Lambda^{A_0}_{11}$ & $\Lambda^{A_0}_{21}$& $B_1$ & $ Z_{m_b} $& $ Z_{m_b}^{\rm TI} $& $ \Lambda^{A_0~{\rm TI}}_{11}$ &$\rho_1$ \\
\hline
3.0 &2 &0.9 &5 & -1.3382 & -0.6789 & -0.3809 &  1.1458 & -0.7854 &  1.0472 &  0.2286\\
4.0 &2 &1.2 &3 & -1.2009 & -0.5833 & -0.3809 &  1.0861 & -0.8508 &  1.0472 &  0.1209\\
6.0 &1 &1.8 &2 & -1.0786 & -0.4599 & -0.3809 &  0.8770 & -0.7854 &  1.0472 &  0.0189\\
8.0 &1 &2.4 &2 & -1.0188 & -0.3635 & -0.3809 &  0.8761 & -0.8508 &  1.0472 & -0.0710\\
10.0 &1 &3.0 &2 & -0.9846 & -0.3017 & -0.3809 &  0.8749 & -0.8901 &  1.0472 & -0.1266\\
12.0 &1 &3.6 &2 & -0.9623 & -0.2574 & -0.3809 &  0.8733 & -0.9163 &  1.0472 & -0.1654\\
16.0 &1 &4.8 &1 & -0.9355 & -0.1990 & -0.3809 &  0.8708 & -0.9490 &  1.0472 & -0.2153\\
32.0 &1 &9.6 &1 & -0.8989 & -0.1040 & -0.3809 &  0.8673 & -0.9981 &  1.0472 & -0.2944\\
64.0 &1 &19.2 &1 & -0.8806 & -0.0534 & -0.3809 &  0.8649 & -1.0227 &  1.0472 & -0.3363\\
128.0 &1 &38.4 &1 & -0.8724 & -0.0270 & -0.3809 &  0.8636 & -1.0349 &  1.0472 & -0.3574
\end{tabular}

\end{table}

\begin{table}[hbt]
\caption{$\rho^{A_0}_1$ contribution to the temporal axial current renormalisation.
\label{TABA0rho1}}
\begin{tabular}{c|c|c|c|c|c|c}
$m_b$ &$ \rho^{A_0}_1(\xi = 0.1) $ &$ \rho^{A_0}_1(\xi = 0.2) $ &$ \rho^{A_0}_1(\xi = 0.3) $ &$ \rho^{A_0}_1(\xi = 0.4) $ &$ \rho^{A_0}_1(\xi = 0.5) $ &$ \rho^{A_0}_1(\xi = 1.0) $ \\
\hline
2.000   & - & - & -& 0.195(3)& 0.070(3)& -0.138(3)\\
3.000   & - & -& 0.229(3)& 0.068(3)& -0.025(2)& -0.159(2)\\
4.000   & -& 0.385(3)& 0.121(2)& -0.009(2)& -0.082(2)& -0.169(2)\\
6.000   & -& 0.224(2)& 0.019(2)& -0.075(2)& -0.126(2)& -0.159(2)\\
8.000  & 0.584(2)& 0.094(2)& -0.071(2)& -0.141(2)& -0.173(2)& -0.172(1)\\
10.000  & 0.435(2)& 0.008(2)& -0.127(1)& -0.181(1)& -0.203(1)& -0.180(1)\\
12.000  & 0.326(2)& -0.052(1)& -0.165(1)& -0.208(1)& -0.223(1)& -0.185(1)\\
16.000  & 0.175(2)& -0.130(1)& -0.215(1)& -0.243(1)& -0.247(1)& -0.190(1)
\end{tabular}

\end{table}

\begin{table}[hbt]
\caption{$\rho^{A_0}_2$ contribution to the temporal axial current renormalisation.
\label{TABA0rho2}}
\begin{tabular}{c|c|c|c|c|c|c}
$m_b$ &$ \rho^{A_0}_2(\xi = 0.1) $ &$ \rho^{A_0}_2(\xi = 0.2) $ &$ \rho^{A_0}_2(\xi = 0.3) $ &$ \rho^{A_0}_2(\xi = 0.4) $ &$ \rho^{A_0}_2(\xi = 0.5) $ &$ \rho^{A_0}_2(\xi = 1.0) $ \\
\hline
2.000   & - & - & -& -0.027(5)& -0.112(4)& -0.142(3)\\
3.000   & - & -& -0.041(5)& -0.129(4)& -0.149(3)& -0.158(2)\\
4.000   & -& 0.067(5)& -0.095(4)& -0.134(3)& -0.145(3)& -0.168(2)\\
6.000   & -& -0.020(4)& -0.079(3)& -0.106(3)& -0.118(2)& -0.158(2)\\
8.000  & 0.239(5)& -0.005(3)& -0.052(3)& -0.077(2)& -0.104(2)& -0.170(2)\\
10.000  & 0.182(4)& 0.020(3)& -0.025(2)& -0.063(2)& -0.070(2)& -0.179(1)\\
12.000  & 0.162(4)& 0.043(3)& -0.006(2)& -0.027(2)& -0.066(2)& -0.183(1)\\
16.000  & 0.182(3)& 0.083(2)& 0.038(2)& -0.016(2)& -0.060(2)& -0.192(1)
\end{tabular}

\end{table}

%
%

\begin{table}[hbt]
\caption{Various contributions to the $\rho_1^{V_k}$ spatial vector current renormalisation for $\xi = 0.3$ .
\label{TABrho1Vk_contributions}
}
\begin{tabular}{c|c|c|c|c|c|c|c|c|c|c}
$m_b$ & $n_b$& $m_c$ & $n_c$ & $\half\{Z_{\psi_c} + Z_{\psi_b}\} + \Lambda^{V_k}_{11}$ & $\Lambda^{V_k}_{21}$& $B_1$ & $ Z_{m_b} $& $ Z_{m_b}^{\rm TI} $& $ \Lambda^{V_k~{\rm TI}}_{11}$ & $\rho_1$ \\
\hline3.0 &2 &0.9 &5 & -1.4361 &  0.6402 &  0.2873 &  1.1496 & -0.7854 &  1.0472 & -0.3281\\
4.0 &2 &1.2 &3 & -1.2605 &  0.5574 &  0.2873 &  1.0932 & -0.8508 &  1.0472 & -0.2991\\
6.0 &1 &1.8 &2 & -1.1096 &  0.4418 &  0.2873 &  0.8795 & -0.7854 &  1.0472 & -0.1862\\
8.0 &1 &2.4 &2 & -1.0372 &  0.3520 &  0.2873 &  0.8786 & -0.8508 &  1.0472 & -0.1024\\
10.0 &1 &3.0 &2 & -0.9971 &  0.2933 &  0.2873 &  0.8748 & -0.8901 &  1.0472 & -0.0407\\
12.0 &1 &3.6 &2 & -0.9719 &  0.2510 &  0.2873 &  0.8734 & -0.9163 &  1.0472 &  0.0039\\
16.0 &1 &4.8 &1 & -0.9406 &  0.1951 &  0.2873 &  0.8720 & -0.9490 &  1.0472 &  0.0626\\
32.0 &1 &9.6 &1 & -0.8983 &  0.1031 &  0.2873 &  0.8696 & -0.9981 &  1.0472 &  0.1638\\
64.0 &1 &19.2 &1 & -0.8806 &  0.0527 &  0.2873 &  0.8663 & -1.0227 &  1.0472 &  0.2243\\
128.0 &1 &38.4 &1 & -0.8721 &  0.0268 &  0.2873 &  0.8649 & -1.0349 &  1.0472 &  0.2555
\end{tabular}

\end{table}
\begin{table}[hbt]
\caption{Various contributions to the $\rho_3^{V_k}$ spatial vector current renormalisation for $\xi = 0.3$ .
\label{TABrho3Vk_contributions}
}
\begin{tabular}{c|c|c|c|c|c|c|c}
$m_b$ & $n_b$& $m_c$ & $n_c$ & $\Lambda^{V_k}_{13}$ & $\Lambda^{V_k}_{23}$ &  $B_3$ & $\rho_3$ \\
\hline3.0 &2 &0.9 &5 & -0.0091 & -0.6212 & -0.1467 &  0.4836\\
4.0 &2 &1.2 &3 & -0.0052 & -0.5416 & -0.1467 &  0.4001\\
6.0 &1 &1.8 &2 & -0.0030 & -0.4333 & -0.1467 &  0.2896\\
8.0 &1 &2.4 &2 & -0.0017 & -0.3472 & -0.1467 &  0.2022\\
10.0 &1 &3.0 &2 & -0.0010 & -0.2897 & -0.1467 &  0.1439\\
12.0 &1 &3.6 &2 & -0.0007 & -0.2483 & -0.1467 &  0.1022\\
16.0 &1 &4.8 &1 & -0.0004 & -0.1933 & -0.1467 &  0.0470\\
32.0 &1 &9.6 &1 & -0.0001 & -0.1024 & -0.1467 & -0.0442\\
64.0 &1 &19.2 &1 & -0.0000 & -0.0527 & -0.1467 & -0.0940\\
128.0 &1 &38.4 &1 & -0.0000 & -0.0267 & -0.1467 & -0.1200
\end{tabular}

\end{table}

\begin{table}[hbt]
\caption{$\rho^{V_k}_1$ contribution to spatial vector current renormalisation.
\label{TABVkrho1}}
\begin{tabular}{c|c|c|c|c|c|c}
$m_b$ &$ \rho^{V_k}_1(\xi = 0.1) $ &$ \rho^{V_k}_1(\xi = 0.2) $ &$ \rho^{V_k}_1(\xi = 0.3) $ &$ \rho^{V_k}_1(\xi = 0.4) $ &$ \rho^{V_k}_1(\xi = 0.5) $ &$ \rho^{V_k}_1(\xi = 1.0) $ \\
\hline
2.000   & - & - & -& -0.364(3)& -0.396(3)& -0.339(3)\\
3.000   & - & -& -0.328(3)& -0.346(2)& -0.329(2)& -0.225(2)\\
4.000   & -& -0.247(3)& -0.299(2)& -0.279(2)& -0.250(2)& -0.149(2)\\
6.000   & -& -0.214(2)& -0.186(2)& -0.151(2)& -0.121(2)& -0.034(2)\\
8.000  & -0.102(2)& -0.142(2)& -0.102(2)& -0.073(2)& -0.050(2)& 0.018(2)\\
10.000  & -0.095(2)& -0.077(2)& -0.041(2)& -0.016(1)& 0.000(1)& 0.055(1)\\
12.000  & -0.062(2)& -0.025(2)& 0.004(1)& 0.022(1)& 0.035(1)& 0.079(1)\\
16.000  & 0.009(2)& 0.047(1)& 0.063(1)& 0.072(1)& 0.079(1)& 0.110(1)
\end{tabular}

\end{table}
\begin{table}[hbt]
\caption{$\rho^{V_k}_2$ contribution to spatial vector current renormalisation.
\label{TABVkrho2}}
\begin{tabular}{c|c|c|c|c|c|c}
$m_b$ &$ \rho^{V_k}_2(\xi = 0.1) $ &$ \rho^{V_k}_2(\xi = 0.2) $ &$ \rho^{V_k}_2(\xi = 0.3) $ &$ \rho^{V_k}_2(\xi = 0.4) $ &$ \rho^{V_k}_2(\xi = 0.5) $ &$ \rho^{V_k}_2(\xi = 1.0) $ \\
\hline
2.000   & - & - & -& -0.168(5)& -0.266(4)& -0.328(3)\\
3.000   & - & -& -0.148(4)& -0.240(4)& -0.250(3)& -0.216(2)\\
4.000   & -& -0.026(5)& -0.181(4)& -0.208(3)& -0.200(3)& -0.139(2)\\
6.000   & -& -0.089(4)& -0.119(3)& -0.113(2)& -0.097(2)& -0.029(2)\\
8.000  & 0.145(5)& -0.056(3)& -0.055(2)& -0.040(2)& -0.031(2)& 0.026(1)\\
10.000  & 0.090(4)& -0.009(3)& -0.002(2)& 0.006(2)& 0.031(2)& 0.060(1)\\
12.000  & 0.079(4)& 0.029(2)& 0.037(2)& 0.058(2)& 0.063(2)& 0.083(1)\\
16.000  & 0.109(3)& 0.093(2)& 0.107(2)& 0.105(2)& 0.104(2)& 0.111(1)
\end{tabular}

\end{table}
\begin{table}[hbt]
\caption{$\rho^{V_k}_3$ contribution to spatial vector current renormalisation.
\label{TABVkrho3}}
\begin{tabular}{c|c|c|c|c|c|c}
$m_b$ &$ \rho^{V_k}_3(\xi = 0.1) $ &$ \rho^{V_k}_3(\xi = 0.2) $ &$ \rho^{V_k}_3(\xi = 0.3) $ &$ \rho^{V_k}_3(\xi = 0.4) $ &$ \rho^{V_k}_3(\xi = 0.5) $ &$ \rho^{V_k}_3(\xi = 1.0) $ \\
\hline
2.000   & - & - & -& 0.487(2)& 0.418(2)& 0.2193(9)\\
3.000   & - & -& 0.484(2)& 0.386(1)& 0.315(1)& 0.1435(7)\\
4.000   & -& 0.545(2)& 0.400(1)& 0.305(1)& 0.2393(9)& 0.0955(5)\\
6.000   & -& 0.433(1)& 0.290(1)& 0.2045(8)& 0.1519(7)& 0.0441(4)\\
8.000  & 0.592(2)& 0.329(1)& 0.2022(8)& 0.1340(7)& 0.0927(6)& 0.0105(3)\\
10.000  & 0.510(1)& 0.2526(9)& 0.1439(7)& 0.0875(6)& 0.0529(5)& -0.0109(2)\\
12.000  & 0.440(1)& 0.1970(8)& 0.1022(6)& 0.0540(5)& 0.0255(4)& -0.0256(2)\\
16.000  & 0.329(1)& 0.1214(6)& 0.0470(4)& 0.0104(4)& -0.0102(3)& -0.0446(2)
\end{tabular}

\end{table}
\begin{table}[hbt]
\caption{$\rho^{V_k}_4$ contribution to spatial vector current renormalisation.
\label{TABVkrho4}}
\begin{tabular}{c|c|c|c|c|c|c}
$m_b$ &$ \rho^{V_k}_4(\xi = 0.1) $ &$ \rho^{V_k}_4(\xi = 0.2) $ &$ \rho^{V_k}_4(\xi = 0.3) $ &$ \rho^{V_k}_4(\xi = 0.4) $ &$ \rho^{V_k}_4(\xi = 0.5) $ &$ \rho^{V_k}_4(\xi = 1.0) $ \\
\hline
2.000   & - & - & -& 0.1458(7)& 0.1664(8)& 0.2188(9)\\
3.000   & - & -& 0.0997(5)& 0.1171(6)& 0.1244(6)& 0.1429(7)\\
4.000   & -& 0.0661(4)& 0.0823(4)& 0.0930(5)& 0.0952(5)& 0.0946(5)\\
6.000   & -& 0.0478(3)& 0.0532(3)& 0.0517(3)& 0.0501(4)& 0.0438(4)\\
8.000  & 0.0259(2)& 0.0339(3)& 0.0310(3)& 0.0271(3)& 0.0234(3)& 0.0100(3)\\
10.000  & 0.0202(2)& 0.0212(2)& 0.0159(2)& 0.0106(2)& 0.0089(2)& -0.0112(2)\\
12.000  & 0.0157(2)& 0.0116(2)& 0.0048(2)& 0.0014(2)& -0.0047(2)& -0.0258(2)\\
16.000  & 0.0081(1)& -0.0015(1)& -0.0080(2)& -0.0158(2)& -0.0224(2)& -0.0448(2)
\end{tabular}

\end{table}

\begin{table}[hbt]
\caption{The overall Lepage-Mackenzie scale $a q^\ast_{V_0}$ for the
current $V_0$ as a function of $M_b$ and $\xi$.
\label{TABqastV0}}
\begin{tabular}{c|c|c|c|c|c|c}
$m_b$ &$ a q^\ast_{V_0}(\xi = 0.1) $ &$ a q^\ast_{V_0}(\xi = 0.2) $ &$ a q^\ast_{V_0}(\xi = 0.3) $ &$ a q^\ast_{V_0}(\xi = 0.4) $ &$ a q^\ast_{V_0}(\xi = 0.5) $ &$ a q^\ast_{V_0}(\xi = 1.0) $ \\
\hline
2.000   & - & - & -& 1.6& 1.6&$ \infty$\\
3.000   & - & -& 1.7& 1.7& 1.8&$ \infty$\\
4.000   & -& 1.7& 1.8& 2.0& 2.2&$ \infty$\\
6.000   & -& 2.1& 2.4& 2.8& 3.1&$ \infty$\\
8.000  & 2.0& 2.6& 3.2& 3.8& 4.3&$ \infty$\\
10.000  & 2.3& 3.3& 4.1& 4.9& 5.5&$ \infty$\\
12.000  & 2.7& 4.0& 5.1& 6.0& 6.8&$ \infty$\\
16.000  & 3.6& 5.5& 7.0& 8.3& 9.5&$ \infty$
\end{tabular}

\end{table}

\begin{table}[hbt]
\caption{The overall Lepage-Mackenzie scale $a q^\ast_{A_k}$ for the
current $A_k$ as a function of $M_b$ and $\xi$.
\label{TABqastAk}}
\begin{tabular}{c|c|c|c|c|c|c}
$m_b$ &$ a q^\ast_{A_k}(\xi = 0.1) $ &$ a q^\ast_{A_k}(\xi = 0.2) $ &$ a q^\ast_{A_k}(\xi = 0.3) $ &$ a q^\ast_{A_k}(\xi = 0.4) $ &$ a q^\ast_{A_k}(\xi = 0.5) $ &$ a q^\ast_{A_k}(\xi = 1.0) $ \\
\hline
2.000   & - & - & -& 3.6& {\bf 6.8}& {\bf 0.4}\\
3.000   & - & -& 3.1& {\bf 27.9}&{\bf 0.2}& 1.6\\
4.000   & -& 2.1& {\bf 5.1}& {\bf 0.8}& 1.4& 2.2\\
6.000   & -& 1.8& {\bf 2.1}& 2.3& 2.5& 3.4\\
8.000  & 1.6& {\bf 0.5}& 3.1& 3.1& 3.3& 4.4\\
10.000  & 1.6& {\bf 24.0}& 3.7& 3.8& 4.1& 5.5\\
12.000  & 1.6& 8.3& 4.2& 4.4& 4.8& 6.5\\
16.000  & 1.7& 6.8& 5.3& 5.7& 6.2& 8.5
\end{tabular}

\end{table}

\begin{table}[hbt]
\caption{Here we show the constituent 
lattice and continuum pieces of $a q^\ast_{A_k}$ in the regions noted in bold in table~\ref{TABqastAk}.
It can be seen that the cancellation between the lattice and continuum one-loop
correction is resulting in unstable behaviour for the overall scale.
\label{TABqastAkcontrib}}
\begin{tabular}{c|c|c|c|c|c|c}
$m_b$ & $m_c$ & $a q^\ast_{LAT}$ & $a q^\ast_{cont}$ & $ Z_{LAT} $ & $Z_{cont}$ & $a q^\ast_{all}$ \\
\hline
2.0 & 1.0 &   1.73 &   0.74 &   0.3039 &  -0.1869 & 6.8\\
2.0 & 2.0 &   1.63 &   1.04 &   0.1453 &  -0.2122 & 0.4\\
3.0 & 1.2 &   1.66 &   0.99 &   0.1992 &  -0.1683 & 27.9\\
3.0 & 1.5 &   1.62 &   1.10 &   0.1501 &  -0.1869 & 0.2\\
4.0 & 1.2 &   1.63 &   1.14 &   0.1795 &  -0.1371 & 5.1\\
4.0 & 1.6 &   1.58 &   1.32 &   0.1200 &  -0.1683 & 0.8\\
6.0 & 1.8 &   1.50 &   1.71 &   0.0878 &  -0.1371 & 2.1\\
8.0 & 1.6 &   1.48 &   1.86 &   0.0971 &  -0.0804 & 0.5\\
10.0 & 2.0 &   1.40 &   2.33 &   0.0660 &  -0.0804 & 24.0\\
\end{tabular}

\end{table}

\end{document}